\documentclass[a4paper,12pt]{article}
\usepackage{amssymb}
\usepackage{amsthm}
\usepackage{color}
\usepackage{amsmath}
\usepackage{graphicx}
\usepackage{natbib}
\usepackage{dsfont}
\usepackage{setspace}

\newtheorem{theo}{Theorem}[section]
\newtheorem{prop}[theo]{Proposition}
\newtheorem{coll}[theo]{Corollary}
\newtheorem{lem}[theo]{Lemma}

\bibpunct[, ]{(}{)}{,}{a}{}{,}

\def\1{\mathds{1}}

\theoremstyle{remark}
\newtheorem{rem}[theo]{Remark}
\newcommand{\be}{\begin{equation}}
\newcommand{\ee}{\end{equation}}
\newcommand{\by}{\begin{eqnarray*}}
\newcommand{\ey}{\end{eqnarray*}}
\renewcommand{\Pr}{\mathbb{P}}
\renewcommand{\le}{\leqslant}

\renewcommand{\leq}{\leqslant}
\renewcommand{\geq}{\geqslant}
\renewcommand{\P}{\mathbb{P}}
\renewcommand{\hat}{\widehat}

\begin{document}

\title{Optimal Payoffs under State-dependent Preferences}
\author{C. Bernard\thanks{{Corresponding author:} Carole Bernard, University
of Waterloo, 200 University Avenue West, Waterloo, Ontario, N2L3G1, Canada.
(email: \texttt{c3bernar@uwaterloo.ca}). Carole Bernard acknowledges support
from NSERC.},\ F. Moraux\thanks{%
Franck Moraux, Univ.\ Rennes 1, 11 rue Jean Mac\'{e}, 35000 Rennes, France.
(email: \texttt{franck.moraux@univ-rennes1.fr}). Franck Moraux acknowledges
financial supports from CREM (CNRS research center) and IAE of Rennes.}, L. R%
\"{u}schendorf\thanks{%
Ludger R\"{u}schendorf, University of Freiburg, Eckerstra\ss e 1, 79104
Freiburg, Germany. (email: \texttt{ruschen@stochastik.uni-freiburg.de}). }\
\ and S. Vanduffel\thanks{%
Steven Vanduffel, Vrije Universiteit Brussel, Pleinlaan 2, 1050 Bruxelles,
Belgium. (email: \texttt{steven.vanduffel@vub.ac.be}). Steven Vanduffel
acknowledges support from BNP Paribas.} }
\maketitle

\begin{abstract}
Most decision theories, including expected utility theory, rank dependent
utility theory and cumulative prospect theory, assume that investors are
only interested in the distribution of returns and not in the states of the
economy in which income is received. Optimal payoffs have their lowest
outcomes when the economy is in a downturn, and this feature is often at
odds with the needs of many investors. We introduce a framework for
portfolio selection within which state-dependent preferences can be
accommodated. Specifically, we assume that investors care about the
distribution of final wealth \textit{and} its interaction with some
benchmark. In this context, we are able to characterize optimal payoffs in
explicit form. Furthermore, we extend the classical expected utility
optimization problem of Merton to the state-dependent situation. Some
applications in security design are discussed in detail and we also solve
some stochastic extensions of the target probability optimization problem.
\end{abstract}

\bigskip

\vspace{3cm}


\section*{Introduction}

Studies of optimal investment strategies are usually based on the
optimization of an expected utility, a target probability or some other
(increasing) \textit{law-invariant} measure. Assuming that investors have
law-invariant preferences is equivalent to supposing that they care only
about the distribution of returns and not about the states of the economy in
which the returns are received. This is, for example, the case under
expected utility theory, Yaari's dual theory, rank-dependent utility theory,
mean-variance optimization and cumulative prospect theory. Clearly, an
optimal strategy has some distribution of terminal wealth and must be the
cheapest possible strategy that attains this distribution. Otherwise, it is
possible to strictly improve the objective and to contradict its optimality.
Dybvig \citeyear{D} was the first to study strategies that reach a given
return distribution at lowest possible cost. Bernard and Boyle \citeyear{BBV}
call these strategies cost-efficient and their properties have been examined
further in Bernard, Boyle and Vanduffel \citeyear{BBV2}. In a fairly general
market setting these authors show that the cheapest way to generate a given
distribution is obtained by a contract whose payoff is decreasing in the
pricing kernel (see also Carlier and Dana \citeyear{CD}). The basic
intuition is that investors consume less in states of economic recession
because it is more expensive to insure returns under these conditions. This
feature is also explicit in a Black-Scholes framework, in which optimal
payoffs at time horizon $T$ are shown to be an \textit{increasing} function
of the price of the risky asset (as a representation of the economy) at time 
$T$. In particular, such payoffs are path-independent.

An important issue with respect to the optimization criteria and the
resulting payoffs under most standard frameworks, is that their worst
outcomes are obtained when the market declines. Arguably, this property of
optimal payoffs does not fit with the aspirations of investors, who may seek
protection against declining markets or, more generally, may consider a
benchmark when making investment decisions. In other words, two payoffs with
the same distribution do not necessarily present the same \textquotedblleft
value\textquotedblright\ for a given investor. Bernard and Vanduffel %
\citeyear{BV} show that insurance contracts can usually be substituted by
financial contracts that have the same payoff distribution but are cheaper.
The existence of insurance contracts that provide protection against
specific events shows that these instruments must present more value for an
investor than financial payoffs that lack this feature. This observation
supports the general observation that investors are more inclined to receive
income in a \textquotedblleft crisis\textquotedblright\ (for example when
their property burns down or when the economy is in recession) than under
\textquotedblleft normal\textquotedblright\ conditions.

This paper makes several theoretical contributions to the study of optimal
investment strategies and highlights valuable applications of its findings
in the areas of portfolio management and security design. First, we clarify
the setting under which optimal investment strategies necessarily exhibit
path-independence. These findings complement Cox and Leland %
\citeyear{CL82,CL} and Dybvig's \citeyear{D} seminal results and underscore
the important role of path-independence in traditional optimal portfolio
selection. Thereafter, as our main contribution, we introduce a framework
for portfolio selection that makes it possible to consider the states in
which income is received. More precisely, it is assumed that investors
target some distribution for their terminal wealth and \textit{additionally }%
aim for a certain (desired) interaction with a random benchmark.\footnote{%
The paper draws its inspiration from the last section in Bernard, Boyle and
Vanduffel \citeyear{BBV2}, in which a constrained cost-efficiency problem is
solved when the joint distribution between the wealth and some benchmark is
determined in some specific area (local dependence constraint).} For
example, the investor may want his strategy to be unrelated to the benchmark
when it decreases but to follow this benchmark when it performs well. Using
our framework, we can characterize optimal payoffs explicitly (Theorems \ref%
{main} and \ref{main2}) in this setting. Such explicit characterizations are
derived independently in Theorems 3.1 and 3.3 of Takahashi and Yamamoto %
\citeyear{Tak} but proved only for cases in which there is a countable
number of states\footnote{%
The results of Takahashi and Yamamoto \citeyear{Tak} are stated in a general
market, but the proof of their basic Theorem 3.1 in Appendix A.1 only holds
when the number of states is countable. The proof of their main theorem,
Theorem 3.3 (Appendix A.3.), is based on the same idea as in Theorem 3.1
(see statement A.3 on page 1571) and is thus also valid in the case of
countable states. Their set up also differs from ours in that these authors
assume that stock prices follow diffusion processes, and they derive the
specific form of the state price density process in this setting (page
1561). In this paper, we do not assume that the underlying stock prices are
diffusion processes and hence the state price process does not need to be of
a specific form (see also our final remarks).}. Furthermore, we show that
optimal strategies in this setting become \textit{conditionally} \textit{%
increasing} functions of the terminal value of the underlying risky asset.

A further main contribution in this part of the paper is the extension of
the classical result of portfolio optimization under expected utility (Cox
and Huang \citeyear{CH}). Specifically, we determine the optimal payoff for
an expected utility maximizer under a dependence constraint, reflecting a
desired interaction with the benchmark (Theorem \ref{EUT2}). The proof
builds on isotonic approximations and their properties (Barlow \textit{et al.%
} \citeyear{BBBB}). We also solve two stochastic generalizations of Browne %
\citeyear{Br} and Cvitani\`{c} and Spivak's \citeyear{CS} classical target
optimization problem in the given state-dependent context.

Finally, we show how these theoretical results are useful in security design
and can help to simplify (and improve) payoffs commonly offered in the
financial markets. We show how to substitute highly path-dependent products
by payoffs that depend only on two underlying assets, which we refer to as
\textquotedblleft twins\textquotedblright . This result is illustrated with
an extensive discussion of the optimality of Asian options. We also
construct alternative payoffs with appealing properties.

The paper is organized as follows. Section \ref{BSmod} outlines the setting
of the investment problem under study. In Section \ref{s0}, we restate basic
optimality results for path-independent payoffs for investors with
law-invariant preferences. We also discuss in detail the sufficiency of
path-independent payoffs when allocating wealth. In Section \ref{S0}, we
point out drawbacks of optimal path-independent payoffs and introduce the
concept of state-dependence used in the following sections. We show that
\textquotedblleft twins\textquotedblright , defined as payoffs that depend
only on two underlying asset values, are optimal for state-dependent
preferences. In Section \ref{TWINS2}, we discuss applications to improve
security designs. In particular, we propose several improvements in the
design of geometric Asian options. In Section \ref{TWINS1}, we solve the
standard Merton problem of maximization of expected utility of final wealth
when the investor constrains the interaction of the final wealth with a
given benchmark. In this context, we also generalize the results of Browne %
\citeyear{Br} and Cvitani\`{c} and Spivak \citeyear{CS} with regard to
target probability maximization. Final remarks are presented in Section \ref%
{Final}. Most of the proofs are provided in the Appendix.

\section{Framework and notation\label{BSmod}}

Consider investors with a given finite investment horizon $T$ and no
intermediate consumption. We model the financial market on a filtered
probability space $\left( \Omega ,\mathcal{F},\mathbb{P}\right) ,$ in which $%
\mathbb{P}$ is the real-world probability measure. The market consists of a
bank account $B$ paying a constant risk-free rate $r>0$, so that $B_{0}$
invested in a bank account at time $0$ yields $B_{t}=B_{0}e^{rt}$ at time $%
t. $ Furthermore, there is a risky asset (say, an investment in stock) whose
price process is denoted by $S=\left( S_{t}\right) _{0\leq t\leq T}.$ We
assume that $S_{t}$ $(0<t<T)$ has a continuous distribution $F_{S_{t}}$. The
no-arbitrage price\footnote{%
The payoffs we consider are all tacitly assumed to be square integrable, to
ensure that all expectations mentioned in the paper exist. In particular, $%
c_{0}\left( X_{T}\right) <+\infty $ for any payoff $X_{T}$ considered
throughout this paper.} at time 0 of a payoff $X_{T}$ paid at time $T>0$ is
given by%
\begin{equation}
c_{0}(X_{T})=\mathbb{E}[\xi _{T}X_{T}],  \label{pricing}
\end{equation}%
where $(\xi _{t})_{t}$ is the state-price density process\footnote{%
The process is commonly so designated. However, strictly speaking, it is not
a density that is at issue, but rather the product of a discount factor
(generally strictly less than 1) and the Radon-Nikodym derivative between
the physical measure and the risk-neutral measure.} ensuring that $(\xi
_{t}S_{t})_{t}$ is a martingale. Moreover, based on standard economic
theory, we assume throughout this paper that state prices are decreasing
with asset prices,\footnote{%
See e.g., Cox, Ingersoll and Ross \citeyear{CIR} and Bondarenko %
\citeyear{Bon}, who shows that property (\ref{function}) must hold if the
market does not allow for statistical arbitrage opportunities, where a
statistical arbitrage opportunity is defined as a zero-cost trading strategy
delivering at $T,$ a positive expected payoff unconditionally, and
non-negative expected payoffs conditionally on $\xi _{T}.$} i.e.,%
\begin{equation}
\xi _{t}=g_{t}(S_{t}),\text{ }t\geq 0,  \label{function}
\end{equation}%
where $g_{t}$ is decreasing (in markets where $\mathbb{E}[S_{T}]>S_{0}e^{rT}$%
). There is empirical evidence that this relationship may not hold in
practice, which is called the pricing kernel puzzle (Brown and Jackwerth %
\citeyear{brown2004pricing}, Grith et al. \citeyear{GHK}). Many explanations
have been provided in the literature (Brown and Jackwerth %
\citeyear{brown2004pricing}, Hens and Reichlin \citeyear{hens2013three}),
including state-dependence of preferences (Chabi-Yo et al. %
\citeyear{chabi2008state}). Therefore, \eqref{function} is not consistent
with a market populated by investors with state dependent preferences.
However, we do not tackle the problem of equilibrium and instead study the
situation of a small investor whose state-dependent preferences do not
influence the pricing kernel that is exogenously given in the market. This
is a commonly studied situation since the work of Karatzas et al. %
\citeyear{Ka}.

The functional form (\ref{function}) for $(\xi _{t})_{t}$ allows us to
present our results regarding optimal portfolios using $(S_{t})_{t}$ as a
reference, which is practical. We will explain in Section \ref{Final} how
the results and characterizations of the optimality of a payoff $X_{T}$ are
tied to its (conditional) anti-monotonicity with $\xi _{T}$ and do not
depend on the functional form \eqref{function} per se. Note that assumption (%
\ref{function}) is satisfied by many popular pricing models, including the
CAPM, the consumption-based models and by exponential L\'{e}vy markets in
which the market participants use Esscher pricing (Vanduffel et al. %
\citeyear{VCMS}, Von Hammerstein et al. \citeyear{LUD}). It is also possible
to use a market model in which prices are obtained using the Growth Optimal
portfolio (GOP) as num\'{e}raire (Platen and Heath \citeyear{PlatenHeath2006}%
), as is discussed further in Section \ref{Final}.

The Black--Scholes model can be seen as a special case of this latter
setting. 
Since we will use it to illustrate our theoretical results, we recall here
its main properties. In the Black--Scholes market, under the real
probability $\mathbb{P}$, the price process $\left( S_{t}\right) _{t}$
satisfies 
\begin{equation*}
\frac{dS_{t}}{S_{t}}=\mu dt+\sigma dZ_{t},
\end{equation*}%
with solution $S_{t}=S_{0}\exp \left( \left( \mu -\frac{\sigma ^{2}}{2}%
\right) t+\sigma Z_{t}\right) $. 
Here, $\left( Z_{t}\right) _{t}$ is a standard Brownian motion, $\mu $ ($>r$%
) the drift and $\sigma >0$ the volatility. The distribution (cdf) of $S_{T}$
is given as 
\begin{equation}
F_{S_{T}}(x)=\mathbb{P}(S_{T}\leq x)=\Phi \left( \frac{\ln \left( \frac{x}{%
S_{0}}\right) -(\mu -\frac{\sigma ^{2}}{2})T}{\sigma \sqrt{T}}\right) ,
\label{cdfst}
\end{equation}%
where $\Phi $ is the cdf of a standard normal random variable. In the
Black--Scholes market, the state-price density process $(\xi _{t})_{t}$ is
unique and $\xi _{t}=e^{-rt}e^{-\theta Z_{t}-\frac{\theta ^{2}t}{2}}$ where $%
\theta =\frac{\mu -r}{\sigma }$. Consequently, $\xi _{t}$ can also be
expressed as a decreasing function of the stock price $S_{t}$, 
\begin{equation}
\xi _{t}=\alpha _{t}\left( \frac{S_{t}}{S_{0}}\right) ^{-\beta },
\label{xiTform}
\end{equation}%
where $\alpha _{t}=\exp \left( \frac{\theta }{\sigma }\left( \mu -\frac{%
\sigma ^{2}}{2}\right) t-\left( r+\frac{\theta ^{2}}{2}\right) t\right) ,\
\beta =\frac{\theta }{\sigma }>0$ (because we assume that $\mathbb{E}%
[S_{T}]=S_{0}e^{\mu T}>S_{0}e^{rT}$).


\section{Law-invariant preferences and optimality of path-independent
payoffs \label{s0}}

In this section, it is understood that investors have \textit{law-invariant}
(state-independent) preferences. This means that they are indifferent
between two payoffs having 
the same payoff distribution (under $\mathbb{P}$). In this case, any random
payoff $X_{T}$ (that possibly depends on the path of the underlying asset
price) admits a path-independent alternative with the same price, which is
at least as good for (i.e., desirable in the eyes of) these investors.
Recall that a payoff is \textit{path-independent} if there exists some
function $f$ such that $X_{T}=f(S_{T})$ holds almost surely. Hence,
investors with law-invariant preferences only need to consider
path-independent payoffs when making investment decisions. Under the
additional (typical) assumption that preferences are \textit{increasing},
any path-dependent payoff can be strictly dominated by a path-independent
one that is increasing in the risky asset.\footnote{%
This dominance can easily be implemented in practice, as all
path-independent payoffs can be replicated statistically with European call
and put options as shown e.g., by Carr and Chou \citeyear{CC} and by Breeden
and Litzenberger \citeyear{BL}.}

Note that results in this section are related closely to the original work
of Cox and Leland \citeyear{CL82,CL}, Dybvig \citeyear{D}, Bernard, Boyle
and Vanduffel \citeyear{BBV2} and Carlier and Dana \citeyear{CD}. These
overview results are recalled here to facilitate the exposition of the
extensions that are developed in the following sections.

\subsection{Sufficiency of path-independent Payoffs}

Proposition \ref{pr3b} shows that for any given payoff there exists a
path-independent alternative with the same price that is at least as good
for investors with law-invariant preferences. Thus, such an investor needs
only to consider path-independent payoffs. All other payoffs are indeed
redundant in the sense that they are not needed to optimize the investor's
objective. The proof of Proposition \ref{pr3b} provides an explicit
construction of an equivalent path-independent payoff.

\begin{prop}[Sufficiency of path-independent payoffs]
\label{pr3b} Let $X_{T}$ be a payoff with price $c$ and having a cdf $F$.
Then, there exists at least one path-independent payoff $f(S_{T})$ with
price $c:=c_{0}(f(S_{T}))$ and cdf $F$.\vspace{-5mm}
\end{prop}

The proof of Proposition \ref{pr3b} is provided in Appendix \ref{proofpr3b}%
.\hfill $\Box $

%

Proposition \ref{pr3b}, however, does not conclude that a given
path-dependent payoff can be strictly dominated by a path-independent one.
The following section shows that the dominance becomes strict as soon as
preferences are increasing.

\subsection{Optimality of path-independent payoffs}

Let $F$ be a payoff distribution with (left-continuous) inverse defined as 
\begin{equation}
F^{-1}(p)=\inf \left\{ x\ |\ F(x)\geq p\right\} .  \label{pseudo}
\end{equation}%
The basic result provided here was originally derived by Dybvig \citeyear{D}
and was presented more generally in Bernard, Boyle and Vanduffel %
\citeyear{BBV2}. It shows how to construct a payoff that generates the
distribution $F$ at minimal price. Such payoff is referred to as \textit{%
cost-efficient} by Bernard and Boyle \citeyear{BBV}.

\begin{theo}[Cost optimality of path-independent payoffs]
\label{theo:pr1} Let $F$ be a cdf. The optimization problem 
\begin{equation}
\underset{X_{T}\sim F}{\min }c_{0}\left( X_{T}\right)  \label{CONSTEFF0}
\end{equation}%
has an almost surely unique solution $X_{T}^{\ast }$ that is
path-independent, almost surely increasing in $S_{T}$ and given by 
\begin{equation}
X_{T}^{\ast }=F^{-1}(F_{S_{T}}(S_{T}))  \label{etoile}
\end{equation}
\end{theo}

\noindent This theorem can be seen as an application of the Hoeffding--Fr%
\'{e}chet bounds recalled in Lemma \ref{lem1}, which is presented in the
Appendix. This result implies that investors with \textit{increasing}
law-invariant preferences may restrict their optimization \textit{strictly}
to the set of path-independent payoffs when making investment decisions.%
\footnote{%
Similar optimality results to those in Theorem \ref{theo:pr1} have been
given in the class of admissible claims $X_{T}$ that are smaller than $F$ in
convex order in Dana and Jeanblanc \citeyear{DJ-05} and in Burgert and R\"{u}%
schendorf \citeyear{BR06}.} The payoff \eqref{etoile} is obviously \textit{%
increasing} in $S_{T}$. In fact, this property characterizes cost-efficiency
because of the a.s. uniqueness of the cost-efficient payoff established in
Theorem \ref{theo:pr1}. Consequently, this implies the following corollary.

\begin{coll}[Cost-efficient payoffs]
\label{CEp} A payoff is cost-efficient if and only if it is almost surely
increasing in $S_{T}$.
\end{coll}

Theorem \ref{theo:pr1} also implies that investors with increasing
law-invariant preferences only invest in path-independent payoffs that are
increasing in $S_{T}$. This is consistent with the literature on optimal
investment problems in which optimal payoffs derived using various
techniques always turn out to exhibit this property.

\begin{coll}[Optimal payoffs for increasing law-invariant preferences]
\label{Rk1} For any payoff $Y_{T}$ at price $c$ that is \textit{not} almost
surely increasing in $S_{T}$ there exists a path-independent payoff $Y_{T}^{{%
\ast }}$ at price $c$ that is a strict improvement for any investor with
increasing and law-invariant preferences.
\end{coll}

A possible choice for $Y_{T}^{{\ast }}$ is given by $Y_{T}^{\ast
}:=F^{-1}(F_{S_{T}}(S_{T}))+(c-c_{0}^{{\ast }})e^{rT},$ in which $c_{0}^{{%
\ast }}$ denotes the price of \ref{etoile}. Note that the payoff $%
Y_{T}^{\ast }$ has price $c$ and is almost surely increasing in $S_{T}$. It
consists in investing an amount $c_{0}^{{\ast }}<c$ in the cost-efficient
payoff (also distributed with $F$) and leaving the remaining funds $c-c_{0}^{%
{\ast }}>0$ in the bank account, so that it is a strict improvement of the
payoff $Y_{T}$.


\section{Optimal payoffs under state-dependent preferences\label{S0}.}

Many of the contracts chosen by law-invariant investors 
do not offer protection in times of economic hardship. In fact, due to the
observed monotonicity property with $S_{T},$ the lowest outcomes for an
optimal (thus, cost-efficient) payoff occur when the stock price $S_{T}$
reaches its lowest levels. 
More specifically, denote by $f\left( S_{T}\right) $ a cost-efficient payoff
(with an increasing function $f$) and by $X_{T}$ another payoff such that
both are distributed with $F$ at maturity. Then, $f\left( S_{T}\right) $
delivers low outcomes when $S_{T}$ is low and it holds\footnote{%
We provide here a short proof of \eqref{ineg}. It is clear that the couple $%
(f(S_{T}),\mathds{1}_{S_{T}<a})$ has the same marginal distributions as $%
(X_{T},\mathds{1}_{S_{T}<a}),$ but $\mathbb{E}[f(S_{T})\mathds{1}%
_{S_{T}<a}]\leq \mathbb{E}[X_{T}\mathds{1}_{S_{T}<a}]$ because $f(S_{T})$
and $\mathds{1}_{S_{T}<a}$ are anti-monotonic (from Lemma \ref{lem1}).} for
all $a\geq 0$ that 
\begin{equation}
\mathbb{E}[f(S_{T})|S_{T}<a]\leq \mathbb{E}[X_{T}|S_{T}<a].  \label{ineg}
\end{equation}%
%
%
%
%
%
%
%
%
%
%
%
%
%
%
%
%
%
%
%
%
%
%
%
%
%
%
%
%
%
%
%
%
%
%
%
%
%
%
%
%
%
%
Let $F$ be the distribution of a put option with payoff $%
X_{T}:=(K-S_{T})^{+}=\max (K-S_{T},0)$. 
Bernard, Boyle and Vanduffel \citeyear{BBV2} show that the payoff of the
cheapest strategy with cdf $F$ can be computed as in \eqref{etoile}. It is
given by $X_{T}^{{\ast }}=(K-a\,S_{T}^{-1})^{+}$ with $a:={S_{0}^{2}\exp ({%
2\left( \mu -{\sigma ^{2}}/{2}\right) T})}$ and is a power put option (with
power -1). 
$X_{T}^{{\ast }}$ is the cheapest way to achieve the distribution $F,$
whereas the first \textquotedblleft ordinary\textquotedblright\ put strategy 
$($with payoff $X_{T})$ is actually the most expensive way to do so. These
payoffs interact with $S_{T}$ in fundamentally different ways, as one payoff
is increasing in $S_{T}$ while the other is decreasing in it. A put option
protects the investor against a declining market, in which consumption is
more expensive than is otherwise typical, whereas the cost-efficient
counterpart $X_{T}^{{\ast }}$ provides no protection but rather emphasizes
the effect of a market deterioration on the wealth received.

As mentioned in the introduction, the use of put options and the demand for
insurance (Bernard and Vanduffel \citeyear{BV}) are signals that many
investors care about states of the economy in which income derived from
investment strategies is received. In particular, they may seek strategies
that provide protection against declining markets or, more generally, that
exhibit a desired dependence with some benchmark.

Hence, in the remainder of this paper, we consider investors who exhibit
state-dependent preferences in the sense that they seek a payoff $X_{T}$
with a desired distribution \textit{and} a desired dependence with a
benchmark asset $A_{T}$. In other words, they fix the joint distribution $G$
of the random couple ($X_{T},A_{T}).$ The optimal state-dependent strategy
is the one that solves for 
\begin{equation}
\underset{\left( X_{T},A_{T}\right) \sim G}{\min }c_{0}\left( X_{T}\right) .
\label{STATEEFF}
\end{equation}%
Note that the setting also includes law-invariant preferences as a special
(limiting) case when $A_{T}$ is deterministic. In this case, we effectively
revert to the framework of state-independent preferences that we discussed
in the previous section. In what follows, we consider as benchmark the
underlying risky asset or any other asset in the market, considered at final
or intermediate time(s). Moreover, to ensure that the impact of
state-dependent preferences on the structure of optimal payoffs is clear, we
have organized the rest of the present section along similar lines to those
of Section \ref{s0}.


\begin{rem}
One can use a copula as a device to model the interaction between payoffs
and benchmarks. The joint distribution $G$ of the couple ($X_{T},A_{T})$ can
be written using a copula $C$. From Sklar's theorem, $%
G(x,a)=C(F_{X_{T}}(x),F_{A_{T}}(a))$, where $C$ is a copula (this
representation is unique for continuously distributed random variables). It
is then clear that the determination of optimal strategies in %
\eqref{STATEEFF} can also be formulated as 
\begin{equation}
\underset{{\substack{X_{T}\sim F,\\ \mathcal{C}_{(X_T,A_T)}=C }}}{\min }%
c_{0}\left( X_{T}\right) ,  \label{eq10}
\end{equation}%
where \textquotedblleft $\mathcal{C}_{(X_{T},A_{T})}=C$\textquotedblright\
means that the copula between the payoff $X_{T}$ and the benchmark $A_{T}$
is $C$. In particular, \eqref{eq10} shows that knowledge of the distribution
of $A_{T}$ is not necessary in order to determine optimal state-dependent
strategies.
\end{rem}

\subsection{Sufficiency of twins}

In this paper, any payoff that writes as $f(S_{T},A_{T})$ or $f(S_{T},S_{t})$
is called a \textit{twin}. 
We show first that, in our state-dependent setting, for any payoff there
exists a twin that is at least as good. When also assuming that preferences
are increasing, we find that optimal payoffs write as twins, and we are able
to characterize them explicitly. Conditionally on $A_{T}$, optimal twins are
increasing in the terminal value of the risky asset $S_{T}$.

The following theorems show that for any given payoff there is a twin that
is at least as good for investors with state-dependent preferences.

\begin{theo}
\label{main} \textup{(Twins as payoffs with a given joint distribution with
a benchmark\linebreak $A_{T}$ and price $c$)\textbf{.}} Let $X_{T}$ be a
payoff with price $c$ having joint distribution $G$ with some benchmark $%
A_{T}$, where $(S_{T},A_{T})$ is assumed to have a joint density with
respect to the Lebesgue measure. Then, there exists at least one twin $%
f\left( S_{T},A_{T}\right) $ with price $c=c_{0}\left( f\left(
S_{T},A_{T}\right) \right) $ having the same joint distribution $G$ with $%
A_{T}$.
\end{theo}

Theorem \ref{main} does not cover the case in which $S_{T}$ plays the role
of the benchmark (because $(S_{T},S_{T})$ has no density). This interesting
case is considered in the following theorem (Theorem \ref{twin2}).

\begin{theo}[Twins as payoffs with a given joint distribution with $S_{T}$
and price $c$]
\label{twin2} Let $X_{T}$ be a payoff with price $c$ having joint
distribution $G$ with the benchmark $S_{T}$. Assume that $(S_{T},S_{t})$ for
some $0<t<T$ has a joint density with respect to the Lebesgue measure. Then,
there exists at least one twin $f(S_{t},S_{T})$ with price $c=c_{0}\left(
f\left( S_{t},S_{T}\right) \right) $ having a joint distribution $G$ with $%
S_{T}$. An example is given by 
\begin{equation}
f(S_{t},S_{T}):=F_{X_{T}|S_{T}}^{-1}(F_{S_{t}|S_{T}}(S_{t})).
\label{twinprop32}
\end{equation}
\end{theo}

\noindent The proofs for Theorems \ref{main} and \ref{twin2} are in Appendix %
\ref{proofmain} and \ref{prooftwin2}. \hfill$\Box$ 

Theorems \ref{main} and \ref{twin2} imply that investors who care about the 
\textit{joint distribution of terminal wealth with some benchmark $A_{T}$}
need only consider the twins in both cases, i.e., when $(A_{T},S_{T})$ is
continuously distributed, as in Theorem \ref{main}, or when $A_{T}$ is equal
to $S_{T},$ as in Theorem \ref{twin2}. These results extend Proposition \ref%
{pr3b} to the presence of a benchmark and state-dependent preferences. All
other payoffs are useless in the sense that they are not needed for these
investors per se.\footnote{%
This finding is consistent with the result obtained by Takahashi and
Yamamoto \citeyear{Tak}, who apply it to replicate a joint distribution in
the hedge fund industry.}

Note that in Theorem \ref{twin2}, $t$ can be chosen freely in $(0,T)$ and
the dependence with respect to $S_{t}$ is not fixed. So, for instance,
replacing $F_{S_{t}}(S_{t})$ with $1-F_{S_{t}}(S_{t})$ in \eqref{twinprop32}
would also lead to the appropriate properties. Hence, there is an infinite
number of twins $f(S_{t},S_{T})$ having the joint distribution $G$ with $%
S_{T}.$ All of them have the same price.\footnote{%
To see this, recall that the joint distribution between the twin $%
f(S_{t},S_{T})$ and $S_{T}$ is fixed and thus also the joint distribution
between the twin and $\xi _{T}$ (as $\xi _{T}$ is a decreasing function of $%
S_{T}$ due to \eqref{function}). All twins $f(S_{t},S_{T})$ with such a
property have the same price $\mathbb{E}[\xi _{T}f(S_{t},S_{T})].$} The
question then arises: how does one select one among them. A natural
possibility is to determine the optimal twin $X_{T}=f(S_{t},S_{T})$ by
imposing an additional criterion. For example, one could define the best
twin $X_{T}$ as the one that minimizes 
\begin{equation}
\mathbb{E}\left[ (X_{T}-H_{T})^{2}\right] ,  \label{min}
\end{equation}%
where $H_{T}$ is another payoff that is not a function of $S_{T}$. This
approach appears natural in the context of simplifying the design of
contracts. For instance, start with a geometric Asian option and compute its
joint distribution $G$ with $S_{T}$. Then, all twins as in \eqref{twinprop32}
have the same price but one of them may be closer to the original Asian
derivative (in the sense of minimizing the distance, as in \eqref{min}).
Note that since all marginal distributions are fixed, the criterion (\ref%
{min}) is equivalent to maximizing the correlation between $X_{T}$ and $%
H_{T} $. We use this criterion in one of our applications (see Section \ref%
{GAS}).

\subsection{Optimality of twins}

Next, we investigate the cost optimality of twins. As discussed above, if
the benchmark $A_{T}$ coincides with $S_{T}$, then all twins that satisfy $%
\left( X_{T},A_{T}\right) \sim G$ have the same cost and the problem of
searching for the cheapest one is not meaningful. However, this observation
is no longer true when the benchmark $A_{T}$ has a density with $S_{T}$. In
this case, the cheapest twin is determined by Theorem \ref{main2} that
extends Theorem \ref{theo:pr1} to the state-dependent case. Theorem \ref%
{theo:pr1} finds that among the infinite number of payoffs with a given
distribution $F$, the cheapest one is increasing in $S_{T}$. In the
state-dependent setting one has that optimal payoffs are increasing in $%
S_{T},$ conditionally on $A_{T}.$

\begin{theo}[Cost optimality of twins]
\label{main2} Assume that $(S_{T},A_{T})$ has joint density 
with respect to the Lebesgue measure. Let $G$ be a bivariate cumulative
distribution function. The optimal state-dependent strategy determined by 
\begin{equation}
\underset{\left( X_{T},A_{T}\right) \sim G}{\min }c_{0}\left( X_{T}\right)
\label{CONSTEFF}
\end{equation}%
has an almost surely unique solution $X_{T}^{{\ast }}$ which is a twin of
the form $f(S_{T},A_{T})$. $X_{T}^{\ast }$ is almost surely increasing in $%
S_{T}$, conditionally on $A_{T},$ and given by 
\begin{equation}
X_{T}^{{\ast }}:=F_{X_{T}|A_{T}}^{-1}(F_{S_{T}|A_{T}}(S_{T})).
\label{xtstar}
\end{equation}
\end{theo}

\noindent The proof of Theorem \ref{main2} is provided in Appendix \ref%
{proofmain2}.\hfill $\Box $

Recall from Section \ref{s0} that when preferences are law-invariant,
optimal payoffs are path-independent and increasing in $S_{T}.$ When
preferences are state-dependent, we observe from expression (\ref{xtstar})
that optimal state-dependent payoffs may become path-dependent, and are
increasing in $S_{T},$ \textit{conditionally} on $A_{T}$. We end this
section with a corollary derived from Theorem \ref{main2}. The result echoes
the one established for investors with law-invariant preferences in the
previous section (Corollary \ref{Rk1})

\begin{coll}[Cheapest twin]
\label{coro2} Assume that $(S_{T},A_{T})$ has joint density 
with respect to the Lebesgue measure. Let $G$ be a bivariate cumulative
distribution function. Let $X_{T}$ be a payoff such that $\left(
X_{T},A_{T}\right) \sim G$. Then, $X_{T}$ is the cheapest payoff if and only
if, conditionally on $A_{T}$, $X_{T}$ is (almost surely) increasing in $%
S_{T} $.
\end{coll}

\noindent The proof of Corollary \ref{coro2} is provided in Appendix \ref%
{proofcoro2}.\hfill $\Box $

\section{Improving security design\label{TWINS2}}

In this section, we show that the results above are useful in designing
balanced and transparent investment policies for retail investors as well as
financial institutions:

\begin{enumerate}
\item If the investor who buys the financial contract has law-invariant
preferences and if the contract is not increasing in $S_{T},$ then there
exists a strictly cheaper derivative (cost-efficient contract) that is
strictly better for this investor. We find its design by applying Theorem %
\ref{theo:pr1}.

\item If the investor buys the contract because of the interaction with the
market asset $S_{T}$, and the contract depends on another asset, then we can
apply Theorem \ref{twin2} to simplify its design while keeping it
\textquotedblleft at least as good.\textquotedblright\ The contract then
depends, for example, on $S_{T}$ and $S_{t}$ for some $t\in (0,T)$.

\item If the investor buys the contract because he likes the dependence with
a benchmark $A_{T},$ which is not $S_{T}$, and if the contract does not only
depend on $A_{T}$ and $S_{T}$, then we use Theorem \ref{main} to construct a
simpler one that is \textquotedblleft at least as good\textquotedblright and
that writes as a function of $S_{T}$ and $A_{T}$. Finally, if the obtained
contract is not increasing in $S_{T}$ conditionally on $A_{T},$ then it is
also possible to construct a strictly cheaper alternative using Theorem \ref%
{main2} and Corollary \ref{coro2}.
\end{enumerate}

We now use the Black--Scholes market to illustrate these three situations.
We begin with the example of an Asian option with fixed strike, followed by
the example of one with floating strike.

\subsection{The geometric Asian twin with fixed strike\label{GAS}}

Consider a fixed strike (continuously monitored) geometric Asian call with
payoff given by 
\begin{equation}
Y_{T}:=\left( G_{T}-K\right) ^{+}.  \label{PAYOFF}
\end{equation}%
Here, $K$ denotes the fixed strike and $G_{T}$ is the geometric average of
stock prices from $0$ to $T,$ defined as 
\begin{equation}
\ln (G_{T}):=\frac{1}{T}\int_{0}^{T}\ln \left( S_{s}\right) ds.  \label{GT}
\end{equation}%
We can now apply the results derived above to design products that improve
upon $Y_{T}$.

\paragraph{Use of cost-efficiency payoff for investors with increasing
law-invariant preferences.}

By applying Theorem \ref{theo:pr1} to the payoff $Y_{T}$ \eqref{PAYOFF}, one
finds that the cost-efficient payoff associated with a fixed strike
(continuously monitored) geometric Asian call is%
\begin{equation}
Y_{T}^{{\ast }}=d\left( S_{T}^{1/\sqrt{3}}-\frac{K}{d}\right) ^{+},
\label{ContGeo}
\end{equation}%
where $d=S_{0}^{1-\frac{1}{\sqrt{3}}}e^{\left( \frac{1}{2}-\sqrt{\frac{1}{3}}%
\/\right) \left( \mu -\frac{\sigma ^{2}}{2}\right) T}$. This is also the
payoff of a power call option, with well-known price%
\begin{equation}
c_{0}\left( Y_{T}^{{\ast }}\right) =S_{0}e^{(\frac{1}{\sqrt{3}}-1)rT+(\frac{1%
}{2}-\frac{1}{\sqrt{3}})\mu T-\frac{\sigma ^{2}T}{12}}\Phi (h_{1})-{K}%
e^{-rT}\Phi (h_{2})  \label{cec}
\end{equation}%
where 
\begin{equation*}
h_{1}=\frac{\ln \left( \frac{S_{0}}{K}\right) +(\frac{1}{2}-\frac{1}{\sqrt{3}%
})\mu T+\frac{r}{\sqrt{3}}T+\frac{1}{12}\sigma ^{2}T}{\sigma \sqrt{\frac{T}{{%
3}}}},\quad h_{2}=h_{1}-\sigma \sqrt{\frac{{T}}{{3}}}.
\end{equation*}%
While the above results can also be found in Bernard, Boyle and Vanduffel %
\citeyear{BBV2}, they are worth considering here for the purpose of
comparison with what follows. 
Note that letting $K$ go to zero provides a cost-efficient payoff that is
equivalent to the geometric average $G_{T}$.

\paragraph{A twin that is useful for investors who care about the dependence
with \protect\boldmath$S_{T}$.}

By applying Theorem \ref{twin2} to the payoff $G_{T}$, we can find a twin
payoff $R_{T}(t)=f(S_{t},S_{T})$ such that 
\begin{equation}
\left( S_{T},R_{T}(t)\right) \sim \left( S_{T},G_{T}\right) .  \label{JOINT}
\end{equation}%
By definition, this twin preserves existing dependence between $G_{T}$ and $%
S_{T}$. However, compared to the original contract it is simpler and
\textquotedblleft less\textquotedblright\ path-dependent, as it depends only
on two values of the path of the stock price. Interestingly, the call option
written on $R_{T}(t)$ and the call option written on $G_{T}$ have the same
joint distribution with $S_{T}$. Consequently,%
\begin{equation}
\left( S_{T},(R_{T}(t)-K)^{+}\right) \sim \left( S_{T},\left( G_{T}-K\right)
^{+}\right) .  \label{JOINT2}
\end{equation}%
$\left( R_{T}(t)-K\right) ^{+}$ is therefore a twin equivalent to the fixed
strike geometric Asian call (as in Theorem \ref{twin2}). We can compute $%
R_{T}(t)$ by applying Theorem \ref{twin2}, and we find that 
\begin{equation}
R_{T}(t)=S_{0}^{\frac{1}{2}-\frac{1}{2\sqrt{3}}\sqrt{\frac{T-t}{t}}}S_{t}^{%
\frac{T}{t}\frac{1}{2\sqrt{3}}\sqrt{\frac{t}{T-t}}}S_{T}^{\frac{1}{2}-\frac{1%
}{2\sqrt{3}}\sqrt{\frac{t}{T-t}}},  \label{RT}
\end{equation}%
where $t$ is freely chosen in $(0,T)$. Details on how \eqref{twinprop32}
becomes \eqref{RT} are provided in Appendix \ref{Geom}.\footnote{%
Formula \eqref{RT} is based on the expression \eqref{twinprop32} for a twin
dependent on $S_{t}$ and $S_{T}$. Note that there is no uniqueness. For
example, $1-F_{S_{t}|S_{T}}(S_{t})$ is also independent of $S_{T},$ and we
can thus also consider $H_{T}(t):=$ $%
F_{X_{T}|S_{T}}^{-1}(1-F_{S_{t}|S_{T}}(S_{t}))$ as a suitable twin ($0<t<T$)
satisfying the joint distribution, as in \eqref{JOINT}. In this case, one
obtains $H_{T}(t)=S_{0}^{\frac{1}{2}+\frac{1}{2\sqrt{3}}\sqrt{\frac{T-t}{t}}%
}S_{t}^{-\frac{T}{t}\frac{1}{2\sqrt{3}}\sqrt{\frac{t}{T-t}}}S_{T}^{\frac{1}{2%
}+\frac{1}{2\sqrt{3}}\sqrt{\frac{t}{T-t}}}.$} The equality of joint
distributions exposed in \eqref{JOINT2} implies that the call option written
on $R_{T}(t)$ has the same price as the original fixed strike (continuously
monitored) geometric Asian call \eqref{PAYOFF}. The time$-0$ price of both
contracts is therefore 
\begin{equation}
c_{0}((R_{T}\left( t\right) -K)^{+})=S_{0}e^{-\frac{rT}{2}-\frac{\sigma ^{2}T%
}{12}}\Phi (\tilde{d}_{1})-Ke^{-rT}\Phi (\tilde{d}_{2}),  \label{price2}
\end{equation}%
where $\tilde{d}_{1}=\frac{\ln (S_{0}/K)+rT/2+\sigma ^{2}T/12}{\sigma \sqrt{%
T/3}}$ and $\tilde{d}_{2}=\tilde{d}_{1}-\sigma \sqrt{T/3}$ (see Kemna and
Vorst \citeyear{KV}).

\paragraph{Choosing among twins.}

The construction in Theorem \ref{twin2} depends on $t$. Maximizing the
correlation between $\ln \left( R_{T}(t)\right) $ and $\ln \left(
G_{T}\right) $ is nevertheless a possible way to select a specific $t$. The
covariance between $\ln (R_{T}(t))$ and $\ln (G_{T})$ is provided by%
\begin{equation*}
\mathop{\mathrm{cov}}\left( \ln \left( R_{T}\left( t\right) \right) ,\ln
\left( G_{T}\right) \right) =\frac{\sigma ^{2}}{2}\left( \frac{T}{2}+\frac{%
\sqrt{t}\sqrt{T-t}}{2\sqrt{3}}\right)
\end{equation*}%
and, by construction of $R_{T}\left( t\right) $, the standard deviations of $%
\ln \left( R_{T}\left( t\right) \right) $ and $\ln \left( G_{T}\right) $ are
both equal to $\sigma \sqrt{\frac{T}{3}}$. Maximizing the correlation
coefficient is therefore equivalent to maximizing the covariance, and thus
of $f(t)=\left( T-t\right) t$. This maximum is obtained for $t^{\ast }=\frac{%
T}{2}$, and the maximal correlation $\rho _{\max }$ between $\ln (R_{T}(t))$
and $\ln (G_{T})$ is 
\begin{equation*}
\rho _{\max }=\frac{3}{4}+\frac{\sqrt{3}\sqrt{\left( T-t^{\ast }\right)
t^{\ast }}}{4T}=\frac{3}{4}+\frac{\sqrt{3}}{8}\approx 0.9665,
\end{equation*}%
which shows that the optimal twin is highly correlated to the initial Asian,
while being considerably simpler. Note that both the maximum correlation and
the optimum $R_{T}(\frac{T}{2})$ are robust to changes in market parameters.

\subsection{The geometric Asian twin with floating strike}

Consider now a floating strike (continuously monitored) Asian put option
defined by%
\begin{equation}
Y_{T}=\left( G_{T}-S_{T}\right)^{+}.  \label{PAYOFFASIANFLOAT}
\end{equation}

For increasing law-invariant preferences, Corollary \ref{Rk1} may be used to
find a cheaper contract that depends on $S_{T}$ only. The cheapest contract
with cdf $F_{Y_{T}}$ is known to be $F_{Y_{T}}^{-1}\left( \Phi \left( \frac{%
\ln \left( \frac{S_{T}}{S_{0}}\right) -(\mu -\frac{\sigma ^{2}}{2})T}{\sigma 
\sqrt{T}}\right) \right) $. Notice that $F_{Y_{T}}^{-1}$ can only be
numerically approximated because the distribution of the difference between
two lognormal distributions is unknown.

If investors care about the dependence with $S_{T}$, by applying Theorem \ref%
{twin2}, one can find twins\ $F_{Y_{T}|S_{T}}^{-1}(F_{S_{t}|S_{T}}(S_{t}))$
as functions of $S_{t}$ and $S_{T},$ which are explicitly given as%
\begin{equation}
\left( S_{0}^{\frac{1}{2}-\frac{1}{2\sqrt{3}}\sqrt{\frac{T-t}{t}}}S_{t}^{%
\frac{T}{t}\frac{1}{2\sqrt{3}}\sqrt{\frac{t}{T-t}}}S_{T}^{\frac{1}{2}-\frac{1%
}{2\sqrt{3}}\sqrt{\frac{t}{T-t}}}-S_{T}\right) ^{+}.  \label{twi}
\end{equation}%
Details can be found in Appendix \ref{Geom2}.

Finally, if investors care about the dependence with $G_{T}$, then it is
possible to construct a cheaper twin because the payoff %
\eqref{PAYOFFASIANFLOAT} is not conditionally increasing in $S_{T}.$
Therefore, it can be strictly improved using Theorem \ref{main2}. The reason
is that we can improve the payoff \eqref{PAYOFFASIANFLOAT} by making it
cheaper while maintaining dependence with $G_{T}.$ Hence, we invoke Theorem %
\ref{main2} (expression \ref{xtstar}) to exhibit another payoff $%
X_{T}=F_{Y_{T}|G_{T}}^{-1}\left( F_{S_{T}|G_{T}}(S_{T})\right) $ such that 
\begin{equation*}
(Y_{T},G_{T})\sim (X_{T},G_{T}),
\end{equation*}%
but so that $X_{T}$ is strictly cheaper. After some calculations, we find
that $X_{T}$ writes as 
\begin{equation}
X_{T}=\left( G_{T}-a\frac{G_{T}^{3}}{S_{T}}\right) ^{+},  \label{OPTI}
\end{equation}%
where $a=\frac{e^{\left( \mu -\frac{\sigma ^{2}}{2}\right) \frac{T}{2}}}{%
S_{0}}.$ Details can be found in Appendix \ref{Geom3}.

Finally, one can easily assess the extent to which the twin \eqref{OPTI} is
cheaper than the initial payoff $Y_{T}$. To do so, we recall the price of a
geometric Asian option with floating strike (the no-arbitrage price of $%
Y_{T} $): 
\begin{equation}
c_{0}(Y_{T})=e^{-rT}\mathbb{E}_{\mathbb{Q}}\left( G_{T}-S_{T}\right)
^{+}=S_{0}e^{-\frac{rT}{2}}\left( \Phi \left( f\right) e^{-\frac{\sigma ^{2}T%
}{12}}-e^{\frac{rT}{2}}\Phi \left( f-\sigma \sqrt{\frac{T}{3}}\right)
\right) ,  \label{float}
\end{equation}%
where $f=\frac{\frac{\sigma ^{2}}{12}T-r\frac{T}{2}}{\sigma \sqrt{\frac{T}{3}%
}}$. Similarly, one finds that 
\begin{equation}
c_{0}(X_{T})=e^{-rT}\mathbb{E}_{\mathbb{Q}}\left( G_{T}-a\frac{G_{T}^{3}}{%
S_{T}}\right) ^{+}=S_{0}e^{-\frac{rT}{2}}\left( \Phi \left( d\right) e^{-%
\frac{\sigma ^{2}T}{12}}-e^{\frac{\mu T}{2}}\Phi \left( d-\sigma \sqrt{\frac{%
T}{3}}\right) \right)  \label{pricy}
\end{equation}%
where $d=\frac{\frac{\sigma ^{2}T}{12}-\mu \frac{T}{2}}{\sigma \sqrt{\frac{T%
}{3}}},$ which we need to compare numerically to \eqref{float}. For example,
when $\mu =0.06,$ $r=0.02,$ $\sigma =0.3$ and $T=1,$ one has $%
c_{0}(Y_{T})=6.74$ and $c_{0}(X_{T})=5.86$, indicating that cost savings can
be substantial. Also note the close correspondence between formulas %
\eqref{float} and \eqref{pricy}. The proofs for these formulas are provided
in Appendix \ref{proofpricy}.

\section{Portfolio management \label{TWINS1}}

This section provides several contributions to the field of portfolio
management. We first derive the optimal investment for an expected utility
maximizer who has a constraint on the dependence with a given benchmark.
Next, we revisit optimal strategies for target probability maximizers (see
Browne \citeyear{Br} and Cvitani\`{c} and Spivak \citeyear{CS}), and we
extend this problem in two directions by adding dependence constraints and
by considering a random target. In both cases, we derive analytical
solutions that are given by twins. From now on, we denote by $W_{0}$ the
initial wealth.

\subsection{Expected utility maximization with dependence constraints\label%
{PM EUM}}

The most prominent decision theory used in various fields of economics is
the expected utility theory (EUT) of von Neumann \& Morgenstern %
\citeyear{VNM}. In the expected utility framework investors assign a utility 
$u(x)$ to each possible level of wealth $x.$ Increasing preferences are
equivalent to an increasing utility function $u(\cdot )$. Assuming that $%
u(\cdot )$ is concave is equivalent to assuming that investors are risk
averse in the sense that for a given budget they prefer a sure income over a
random one with the same mean. In their seminal paper on optimal portfolio
selection, Cox and Huang \citeyear{CH} showed how to obtain the optimal
strategy for a risk averse expected utility maximizer; see also Merton %
\citeyear{M} and He and Pearson \citeyear{HP-91a},\citeyear{HP-91b}. We
recall this classical result in the following theorem.

\begin{theo}[Optimal payoff in EUT]
\label{EUT1}Consider a utility function $u(\cdot )$ defined on $(a,b)$ such
that $u(\cdot )$ is continuously differentiable and strictly increasing, $%
u^{\prime }(\cdot )$ is strictly decreasing, $\lim_{x\searrow a}u^{\prime
}(x)=+\infty $ and $\lim_{x\nearrow b}u^{\prime }(x)=0.$ Consider the
following portfolio optimization problem: 
\begin{equation}
\max_{\mathbb{E}[\xi _{T}X_{T}]=W_{0}}\mathbb{E}[u(X_{T})].
\label{opt-prob-EUT1}
\end{equation}%
The optimal solution to this problem is given by 
\begin{equation}
X_{T}^{\ast }=(u^{\prime })^{-1}\left( \lambda \xi _{T}\right) ,
\label{EUT-XTstar}
\end{equation}%
where $\lambda $ is such that $\mathbb{E}\left[ \xi _{T}\left( u^{\prime
}\right) ^{-1}\big(\lambda \xi_T\big)\right] =W_{0}$.
\end{theo}

Note that the optimal EUT payoff $X_{T}^{\ast }$ is decreasing in $\xi _{T}$
and thus increasing in $S_{T}$ (illustration of the results derived in
Section \ref{s0}), which highlights the lack of protection of optimal
portfolios when markets decline. To account for this, we give the investor
the opportunity to maintain a desired dependence with a benchmark portfolio
(e.g., representing the financial market). This extends earlier results on
expected utility maximization with constraints, such as those of Brennan and
Solanki \citeyear{BrSo}, Brennan and Schwartz \citeyear{BrSc}, He and
Pearson \citeyear{HP-91a},\citeyear{
HP-91b}, Basak \citeyear{Ba}, Grossman and Zhou \citeyear{GrZh}, Sorensen %
\citeyear{So} and Jensen and Sorensen \citeyear{JS}. These studies were for
the most part concerned with the expected utility maximization problem when
investors want a lower bound on their optimal wealth either at maturity or
throughout some time interval. When this bound is deterministic, this
corresponds to classical portfolio insurance. Boyle and Tian \citeyear{BT}
extend and unify the various results by allowing the benchmark to be beaten
with some confidence. They consider the following maximization problem over
all payoffs $X_{T}:$ 
\begin{equation}
\underset{\substack{\Pr (X_{T}\geq A_{T})\geq \alpha, \\ c_{0}(X_{T})=W_{0}}}%
{\max }{\mathbb{E}[u(X_{T})],}  \label{BT}
\end{equation}%
where $A_{T}$ is some benchmark (e.g., the portfolio of another manager in
the market). In Theorem 2.1 (page 327) of Boyle and Tian \citeyear{BT}, the
optimal contract ${X_{T}^{{\ast }}}$ is derived explicitly (under some
regularity conditions ensuring feasibility of the stated problem), and it is
an optimal twin.\footnote{%
The observation that in the given context optimal payoffs write as twins is
also consistent with the solutions of the constrained portfolio optimization
problems considered in Bernard, Chen and Vanduffel \citeyear{BCVQF} and
Bernard and Vanduffel \citeyear{BVEJOR}.}

This also follows from our results. Assume that the solution to \eqref{BT}
exists, and denote it by $X_{T}^{{\ast }}$. Then let $G$ be the bivariate
cdf of $(X_{T}^{{\ast }},A_{T})$. The cheapest way to preserve this joint
bivariate cdf is obtained by a twin $f(A_{T},S_{T}),$ which is increasing in 
$S_{T}$ conditionally on $A_{T}$ (see Corollary \ref{coro2}). Hence, $X_{T}^{%
{\ast }}$ must also be of this form, otherwise one can easily contradict the
optimality of $X_{T}^{{\ast }}$ to the problem. Thus, the solution to
optimal expected utility maximization with the additional probability
constraint (when it exists) is an optimal twin. By similar reasoning, this
result also holds when there are several probability constraints involving
the joint distribution of terminal portfolio value $X_{T}$ and benchmark $%
A_{T}$. 

The following theorem extends Theorem \ref{EUT1} and the referenced
literature above by considering an expected utility maximization problem in
which the investor fixes the dependence with a benchmark. Doing so amounts
to specifying up front the joint copula of $(X_{T},A_{T})$. 
Hence, let us assume that the copula between $X_{T}$ and $A_{T}$ is
specified to be $C$, i.e., $\mathcal{C}_{(X_{T},A_{T})}=C.$ We formulate the
following portfolio optimization problem 
\begin{equation}
\underset{\substack{ c_{0}(X_{T})=W_{0} \\ \mathcal{C}_{(X_{T},A_{T})}=C }}{%
\max }\mathbb{E}\left( u(X_{T})\right) .  \label{opt-prob}
\end{equation}%
In order to solve the expected utility optimization problem with dependence
constraints \eqref{opt-prob}, we denote by $C_{1\mid A_{T}}$ the conditional
distribution of the first component, given $A_{T}$ (or equivalently given $%
F_{A_{T}}(A_{T})$) and define 
\begin{equation}
U_{T}=F_{S_{T}|A_{T}}(S_{T})\text{ and }Z_{T}=C_{1|A_{T}}^{-1}(U_{T}).
\label{uz}
\end{equation}%
Note that when $(A_{T},S_{T})$ has a joint density, then $U_{T}$ and $Z_{T}$
are uniformly distributed on $(0,1)$ and $(Z_{T},A_{T})$ has copula $C$ (see
also Lemma \ref{lem2}). Theorem \ref{EUT2} makes also use of the projection
on the convex cone 
\begin{equation}
M_{\downarrow }:=\{f\in L^{2}[0,1];f\text{ decreasing}\},  \label{Md}
\end{equation}%
which is a subset of $L^{2}[0,1]$ equipped with the Lebesgue measure and the
standard $||\cdot ||_{2}$ norm. For an element $\varphi \in L^{2}[0,1]$, we
denote by $\hat{\varphi}=\pi _{M_{\downarrow }}(\varphi )$ the projection of 
$\varphi $ on $M_{\downarrow }$. $\hat{\varphi}$ can be interpreted as the
best approximation of $\varphi $ by a decreasing function for the $||\cdot
||_{2}$ norm.

\begin{theo}[Optimal payoff in EUT with dependence constraint]
\label{EUT2} Consider a utility function $u(\cdot )$ as in Theorem \ref{EUT1}
and assume that $(A_{T},S_{T})$ has a joint density. Let $H_{T}=\mathbb{E}%
(\xi _{T}|Z_{T})=\varphi (Z_{T})$ and $\hat{H}_{T}=\hat{\varphi}(Z_{T})$ in
which $Z_{T}$ is defined as in (\ref{uz}). Then, the solution to the
optimization problem \eqref{opt-prob} is given by 
\begin{equation}
\hat{X}_{T}=\left( u^{\prime }\right) ^{-1}\left( \lambda \hat{H}_{T}\right)
,
\end{equation}%
where $\lambda $ is such that $\mathbb{E}\left[ \xi _{T}\left( u^{\prime
}\right) ^{-1}\left( \lambda \hat{H}_{T}\right) \right] =W_{0}$.
\end{theo}

The proof of Theorem \ref{EUT2} is provided in Appendix \ref{proofEUT2}%
.\hfill $\Box $

\begin{rem}
\label{rem:5-3}In the case that $H_{T}=\mathbb{E}(\xi _{T}|Z_{T})$ is
decreasing in $Z_{T}$, we obtain, as solution to \eqref{opt-prob}, 
\begin{equation}
\hat{X}_{T}=\left( u^{\prime }\right) ^{-1}(\lambda H_{T}).
\label{eq:rem-5-3}
\end{equation}%
In this case, the proof of Theorem \ref{EUT2} can be simplified and reduced
to the classical optimization result in Theorem \ref{EUT1} since by Theorem %
\ref{main2} an optimal solution $X_{T}$ is unique and satisfies 
\begin{equation*}
X_{T}=F_{X_{T}|A_{T}}^{-1}(F_{S_{T}|A_{T}}(S_{T})).
\end{equation*}%
By Lemma \ref{lem2} one can conclude that $X_{T}=F_{X_{T}}^{-1}(Z_{T})$,
i.e., $X_{T}$ is an increasing function of $Z_{T}$. Theorem \ref{EUT1} then
allows one to find the optimal element in this class.
\end{rem}

\bigskip

\begin{rem}
\label{rem:5-4} The determination of the isotonic approximation $\hat{\varphi%
}$ of $\varphi $ is a well-studied problem (see Theorem 1.1 in Barlow 
\textit{et al.} \citeyear{BBBB}). $\hat{\varphi}$ is the slope of the
smallest concave majorant $SCM(\varphi )$ of $\varphi $, i.e., $\hat{\varphi}%
=(SCM(\varphi ))^{\prime }$. In Barlow \textit{et al.} \citeyear{BBBB} the
projection on $M_{\uparrow }$ is given as the slope of the greatest convex
minorant $GCM(\varphi )$ of $\varphi $. Fast algorithms are known to
determine $\hat{\varphi}$.
\end{rem}

\bigskip

\begin{rem}
Some special cases of interest concern the study of the optimum when the
copula constraint is the lower or upper Fr\'{e}chet bound. If in Theorem \ref%
{EUT2} the copula $C$ is the upper Fr\'{e}chet bound, then $%
Z_{T}=F_{A_{T}}(A_{T})$. When $A_{T}=S_{T}$, then $H_{T}=E[\xi
_{T}|A_{T}]=\xi _{T}$ and we find that $\hat{X}_{T}$ is equal to the optimal
portfolio when there is no dependence constraint (Theorem \ref{EUT1}). This
result is intuitive because the dependence constraint that we impose implies
that that the optimum is increasing in $S_{T}$, which is a feature that
arises naturally in the unconstrained problem. If $A_{T}=S_{t}$, then $%
H_{T}=E[\xi _{T}|S_{t}]$ 
is decreasing in $S_{t}$. Thus, $\hat{H}_{T}=H_{T}$ and the optimum can be
explicitly calculated (see also the example below). Finally, if in Theorem %
\ref{EUT2} the copula $C$ is the lower Fr\'{e}chet bound , then $%
Z_{T}=1-F_{A_{T}}(A_{T})$. Assume that $A_{T}=S_{T}$, then $H_{T}=E[\xi
_{T}|Z_{T}]=\xi _{T}$, which is increasing in $S_{T}$ and therefore
decreasing in $Z_{T}$. The isotonic approximation is the constant. Hence,
the optimal portfolio is also a constant, i.e., the budget is entirely
invested in the risk-free asset.%
\end{rem}

\bigskip

\paragraph{Example (CRRA investor)\label{EXAMPLE subsection 5.1}}

Next, we illustrate Theorem \ref{EUT2} by a comparison of the optimal wealth 
$\hat{X}_{T}$ derived under a dependence constraint (Theorem \ref{EUT2})
with the optimal wealth $X_{T}^{\star }$ derived with no constraints on
dependence (Theorem \ref{EUT1}). $W_{0}$ stands for the initial wealth and
we set the benchmark $A_{T}$ equal to $S_{t}$ for some $0<t<T.$ We assume
also that the dependence between $S_{t}$ and the final wealth is described
by a Gaussian copula $C$ with correlation coefficient $\rho \in \left[ -%
\sqrt{1-\frac{t}{T}},1\right) $. Consider a CRRA utility function with risk
aversion $\eta >0:$ 
\begin{equation*}
u\left( x\right) :=\left\{ 
\begin{array}{cc}
\frac{x^{1-\eta }}{1-\eta } & \text{when }\eta \neq 1 \\ 
\ln \left( x\right) & \text{when }\eta =1%
\end{array}%
\right. .
\end{equation*}%
The standard Merton problem (\ref{opt-prob-EUT1}) exposed in Theorem \ref%
{EUT1} involves no dependence constraint on the final wealth. The solution
is $X_{T}^{\star }=\left( u^{\prime }\right) ^{-1}(\lambda \xi _{T})$ where $%
\lambda $ is found to meet the initial wealth constraint ($\mathbb{E}\left[
\xi _{T}X_{T}^{\star }\right] =W_{0}$). It is straightforward to verify that
for all $\eta >0$ the optimal wealth is given by 
\begin{equation}
X_{T}^{\star }\left( \eta \right) =\lambda ^{-\frac{1}{\eta }}\xi _{T}^{-%
\frac{1}{\eta }}=W_{0}e^{rT}e^{-\frac{1}{\eta }\frac{\theta }{\sigma }\left(
\mu -\frac{\sigma ^{2}}{2}\right) T+\left( \frac{1}{\eta }-\frac{1}{2\eta
^{2}}\right) \theta ^{2}T}\left( \frac{S_{T}}{S_{0}}\right) ^{\frac{1}{\eta }%
\frac{\theta }{\sigma }}  \label{XStarCRRA}
\end{equation}%
Observe that the dependence between $X_{T}^{\ast }\left( \eta \right) $ and $%
S_{t}$ is characterized by the Gaussian copula with correlation parameter%
\begin{equation}
\text{corr}\left( \ln (X_{T}^{\ast }\left( \eta \right) \right) ,\ln
(S_{t}))=\sqrt{\frac{t}{T}}.  \label{rholevel}
\end{equation}

When there is a constraint on the dependence, we show in Appendix \ref%
{ProofExample1} that the solution to the optimization problem (\ref{opt-prob}%
) (that is the optimal wealth satisfying the initial budget and the
dependence constraint) is given as 
\begin{eqnarray}
\hat{X}_{T}\left( \eta \right) &=&W_{0}e^{rT}e^{-\frac{1}{\eta }\frac{\theta 
}{\sigma }\left( \mu -\frac{\sigma ^{2}}{2}\right) \left( \rho \sqrt{t}+%
\sqrt{(1-\rho ^{2})(T-t)}\right) ^{2}+\left( \frac{1}{\eta }-\frac{1}{2\eta
^{2}}\right) \theta ^{2}\left( \rho \sqrt{t}+\sqrt{(1-\rho ^{2})(T-t)}%
\right) ^{2}}\times  \notag \\
&&\left( \frac{S_{T}}{S_{0}}\right) ^{\frac{\theta }{\eta \sigma }\left(
\rho \sqrt{t}+\sqrt{(1-\rho ^{2})(T-t)}\right) \frac{\sqrt{1-\rho ^{2}}}{%
\sqrt{T-t}}}\left( \frac{S_{t}}{S_{0}}\right) ^{\frac{\theta }{\eta \sigma }%
\left( \rho \sqrt{t}+\sqrt{(1-\rho ^{2})(T-t)}\right) \left[ \frac{\rho }{%
\sqrt{t}}-\frac{\sqrt{1-\rho ^{2}}}{\sqrt{T-t}}\right] }.  \label{XHatCRRA}
\end{eqnarray}%
Note that the expressions (\ref{XStarCRRA}) and (\ref{XHatCRRA}) coincide
when $\rho =\sqrt{\frac{t}{T}}.$ The basis reason for this feature is that
the unconstrained optimum has correlation $\sqrt{\frac{t}{T}}$ with $S_{t}.$
When $\eta \neq 1,$ the expected utilities of $X_{T}^{\star }\left( \eta
\right) $ and $\hat{X}_{T}\left( \eta \right) $ are given by%
\begin{equation*}
\mathbb{E}\left[ u\left( X_{T}^{\star }\left( \eta \right) \right) \right] =%
\frac{1}{1-\eta }W_{0}^{1-\eta }e^{\left( 1-\eta \right) rT+\frac{1}{2}\frac{%
1-\eta }{\eta }\theta ^{2}T}
\end{equation*}%
and 
\begin{equation*}
\mathbb{E}\left[ u\left( \hat{X}_{T}\left( \eta \right) \right) \right] =%
\frac{1}{1-\eta }W_{0}^{1-\eta }e^{\left( 1-\eta \right) rT+\frac{1}{2}\frac{%
1-\eta }{\eta }\theta ^{2}\left( \rho \sqrt{t}+\sqrt{(1-\rho ^{2})(T-t)}%
\right) ^{2}},
\end{equation*}%
respectively. In the case that $\eta =1$, i.e., the log-utility case $%
u\left( x\right) :=\ln \left( x\right) $, we find that 
\begin{equation*}
\mathbb{E}\left[ u\left( \hat{X}_{T}\right) \right] =\ln \left( W_{0}\right)
+rT+\frac{1}{2}\theta ^{2}\left( \rho \sqrt{t}+\sqrt{1-\rho ^{2}}\sqrt{T-t}%
\right) ^{2}
\end{equation*}%
and 
\begin{equation*}
\mathbb{E}\left[ u\left( X_{T}^{\star }\right) \right] =\ln \left(
W_{0}\right) +rT+\frac{1}{2}\theta ^{2}T,
\end{equation*}%
respectively.

\begin{figure}[tbh]
\begin{tabular}{cc}
\includegraphics[width=7cm,height=7cm]{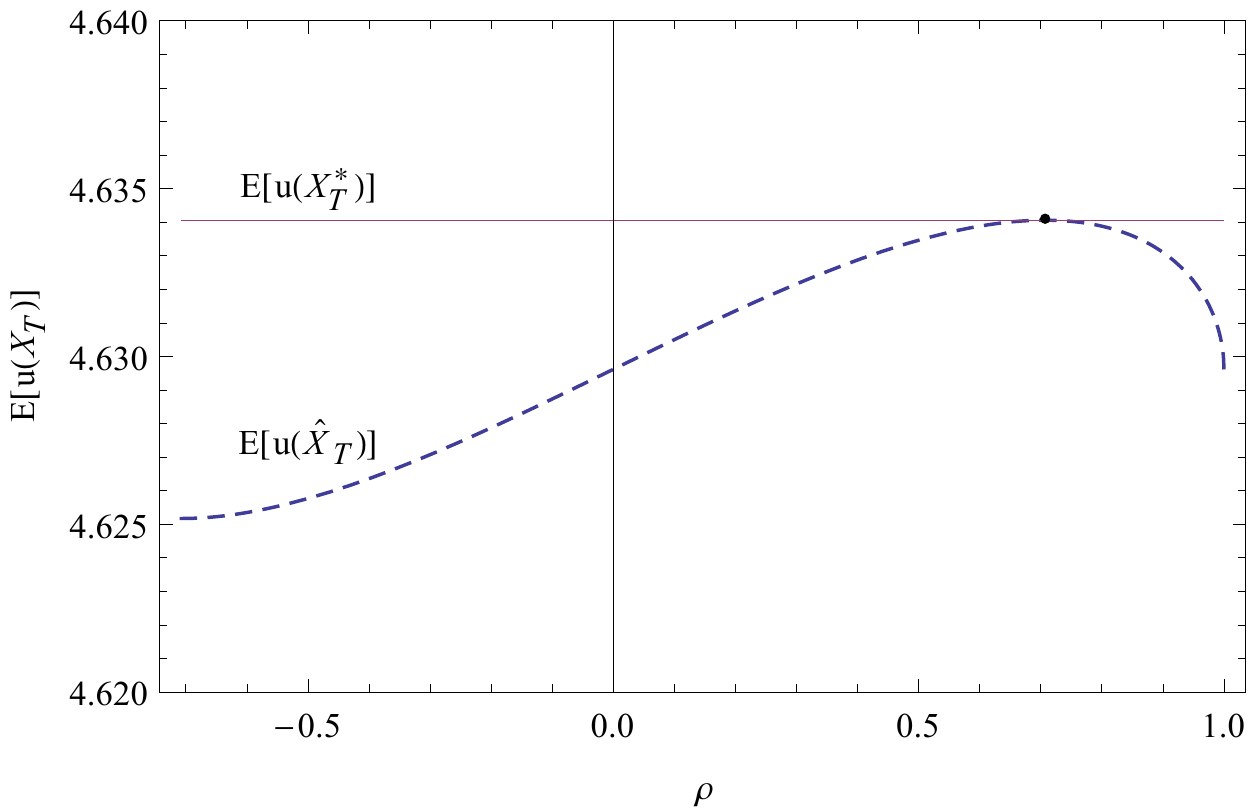} & %
\includegraphics[width=7cm,height=7cm]{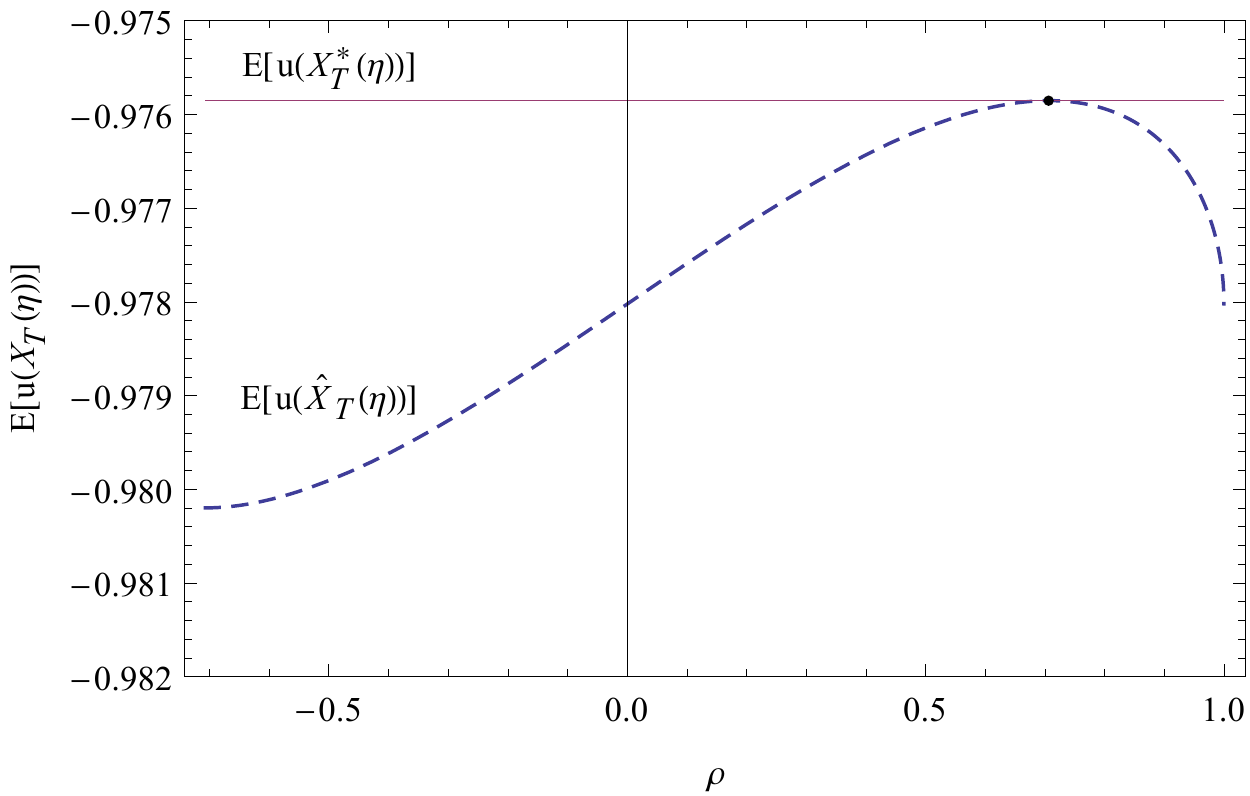} \\ 
$\eta=1 $ ($log$ utility) & $\eta=2$%
\end{tabular}%
\caption{Expected utility as a function of $\protect\rho $ for a CRRA
investor, with and without dependence constraint.}
\label{F1}
\end{figure}

Assume that $t=T/2$ for the numerical application so that $S_{T/2}$ is the
benchmark. Using an initial wealth $W_{0}=100$ and the same set of
parameters as in the previous section, $\mu =0.06,$ $r=0.02,$ $\sigma =0.3$
and $T=1.$ Figure \ref{F1} plots the expected utility as a function of $\rho 
$ for the constrained payoff ($\hat{X}_{T}$) and we have an horizontal line
corresponding to the expected utility of $X_{T}^{\ast }$. Note that they
share exactly one common point corresponding to the level of correlation
found in \eqref{rholevel}.

\subsection{Target probability maximization}

Target probability maximizers are investors who, for a given budget (initial
wealth) and a given time frame, want to maximize the probability that the
final wealth achieves some fixed target $b$. In a Black--Scholes financial
market model, Browne \citeyear{Br} and Cvitani\`{c} and Spivak \citeyear{CS}
derive the optimal investment strategy for these investors using stochastic
control theory and show that it is optimal to purchase a digital option
written on the risky asset. We show that their results follow from Theorem %
\ref{theo:pr1} in a more straightforward way.

\begin{prop}[Browne's original problem]
\label{BrP} Let $W_{0}$ be the initial wealth and let $b>W_{0}e^{rT}$ be the
desired target.\footnote{%
If $b\leq W_{0}e^{rT},$ then the problem is not interesting since an
investment in the risk-free asset allows the investor to reach a 100\%
probability of beating the target $b.$} 
The solution to the following target probability maximization problem, 
\begin{equation}
\underset{X_{T}\geq 0,\text{ }c_{0}(X_{T})=W_{0}}{\max }\P [X_{T}\geq b],
\end{equation}%
is given by the payoff%
\begin{equation}
X_{T}^{\ast }=b\,\mathds{1}_{\{S_{T}>\lambda \}},
\end{equation}%
in which $\lambda $ is given by $b\mathbb{E}\left( \xi _{T}\mathds{1}%
_{\{S_{T}>\lambda \}}\right) =W_{0}$.
\end{prop}

\noindent The proof of this proposition is provided in Appendix \ref{proofBr}%
. In a Black--Scholes market one easily verifies that $\lambda =S_{0}\exp
\left( {(r-\frac{\sigma ^{2}}{2})T-\sigma \sqrt{T}\Phi ^{-1}\left( \frac{%
W_{0}e^{rT}}{b}\right) }\right) $. \hfill $\Box $\bigskip

A target probability maximizing strategy is essentially an all-or-nothing
strategy. Intuitively, investors might not be attracted by the design of the
optimal payoff, which maximizes the probability beating a fixed target. The
obtained wealth depends solely on the ultimate value of the underlying risky
asset, which makes it highly dependent on final market behavior and thus
prone to unexpected and brutal changes. 
Our first extension concerns the case of a stochastic target, so that
preferences become state-dependent.

\begin{theo}[Target probability maximization with a random target]
\label{TO2bis} Let $W_{0}$ be the initial wealth and let $B$ be the random
target such that $(B,S_{T})$ has a density. The solution to the random
target probability maximization problem,%
\begin{equation}
\underset{X_{T}\geq 0,\text{ }c_{0}(X_{T})=W_{0}}{\max }\P [X_{T}\geq B],
\end{equation}%
is given by the payoff 
\begin{equation}
X_{T}^{\ast }=B\mathds{1}_{\left\{ B{\xi _{T}}<\lambda \right\} },
\end{equation}%
in which $\lambda $ is implicitly given by $\mathbb{E}\left[ B\xi _{T}%
\mathds{1}_{\left\{ B\xi _{T}<\lambda \right\} }\right] =W_{0}$.
\end{theo}

\noindent The proof of this proposition is provided in Appendix \ref%
{proofTO2bis}.\hfill $\Box $


Our second extension assumes a fixed dependence with a benchmark in the
financial market. We now consider the problem of an investor who, for a
given budget, aims to maximize the probability that the final wealth will
achieve some fixed target while preserving a certain dependence with a
benchmark.

\begin{theo}[Target probability maximization with a random benchmark]
\label{TO2} Let $W_{0}$ be the initial wealth and let $b>W_{0}e^{rT}$ the
desired target for final wealth. Assume that the pair $(A_{T},S_{T})$ has a
density. Then the solution to the target probability optimization problem
with random benchmark $A_{T}$, 
\begin{equation}
\underset{\substack{ X_{T}\geq 0, c_{0}(X_{T})=W_{0}, \\
\mathcal{C}_{(X_T,A_T)}=C}}{\max }\P [X_{T}\geq b],
\end{equation}%
is given by 
\begin{equation}
X_{T}^{\ast }=b\mathds{1}_{\left\{ Z_{T}>\lambda \right\} },
\end{equation}%
in which $\lambda $ is determined by $b\mathbb{E}\left[ \xi _{T}\mathds{1}%
_{\left\{ Z_{T}>\lambda \right\} }\right] =W_{0}$ and $Z_{T}$ is defined as
in (\ref{uz}).
\end{theo}

\noindent The proof of this result is provided in Appendix \ref{proofTO2}%
.\hfill $\Box $

The result derived in Theorem \ref{TO2} holds in particular when $%
A_{T}=S_{t} $ ($0<t<T$) and when $C$ is a Gaussian copula with correlation
coefficient $\rho .$ Then, the optimal solution is explicit and equal to 
\begin{equation}
X_{T}^{\ast }=b\mathds{1}_{\left\{ S_{t}^{\alpha }S_{T}>\lambda \right\} },
\label{exampleTPM}
\end{equation}%
with $\alpha =\sqrt{\frac{T-t}{t(1-\rho ^{2})}}\rho -1,$ and $\lambda
=S_{0}^{\alpha +1}\exp \left( {(r-\frac{\sigma ^{2}}{2})(\alpha t+T)-\sigma 
\sqrt{k}\Phi ^{-1}\left( \frac{W_{0}e^{rT}}{b}\right) }\right) $ with $%
k=(\alpha +1)^{2}t+(T-t)=\frac{T-t}{1-\rho ^{2}}$. The proof of (\ref%
{exampleTPM}) is provided in Appendix \ref{proofexampleTPM}.

\paragraph{Illustration of target probability maximization}

\begin{figure}[tbh]
\begin{center}
\begin{tabular}{c}
\includegraphics[width=12cm,height=7cm]{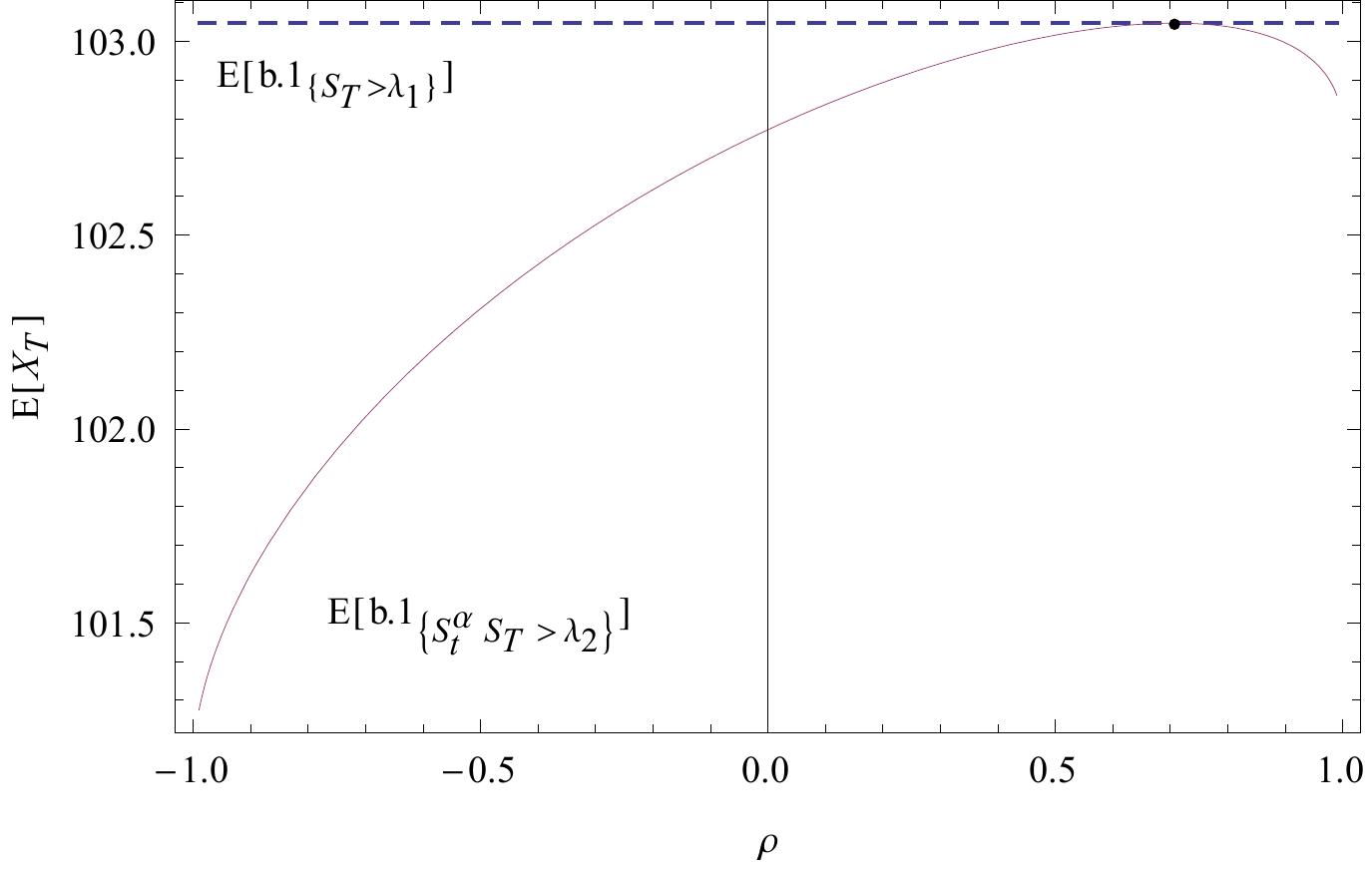}%
\end{tabular}%
\end{center}
\caption{Expected payoff as a function of $\protect\rho $ for the different
target probability maximization strategies considered in Theorem \protect\ref%
{BrP} and Theorem \protect\ref{TO2}.}
\label{F2}
\end{figure}

Let us compare the payoffs that arise from the unconstrained target
probability maximization problem in Theorem \ref{BrP} and the constrained
maximization problem in Theorem \ref{TO2}. We use the same set of parameters
as in Section \ref{PM EUM}, i.e., $\mu =0.06,$ $r=0.02,$ $\sigma =0.3$ and $%
T=1.$ We also take $S_{0}^{{}}=100$ and $b=106$. 
In Figure \ref{F2}, we plot for both payoffs their expected value as a
function of $\rho $. The optimum for the unconstrained target optimization
problem in Theorem \ref{BrP} is given by $b\mathds{1}_{\{S_{T}>\lambda
_{1}\}}$ in which $\lambda _{1}$ is such that the budget constraint is
satisfied. Its expected value is given as 
\begin{equation*}
\mathbb{E}\left[ b\mathds{1}_{\left\{ S_{T}>\lambda _{1}\right\} }\right] =b{%
\Phi }\left[ \theta \sqrt{T}+{\Phi }^{-1}\left[ \frac{W_{0}e^{rT}}{b}\right] %
\right] .
\end{equation*}%
By similar reasoning, we find for the expected value of the optimum of
Theorem \ref{TO2},
\begin{equation*}
\mathbb{E}\left[ b\mathds{1}_{\left\{ S_{t}^{\alpha }S_{T}>\lambda
_{2}\right\} }\right] =b{\Phi }\left[ \theta \frac{\alpha t+T}{\sqrt{k}}+{%
\Phi }^{-1}\left[ \frac{W_{0}e^{rT}}{b}\right] \right] ,
\end{equation*}%
in which $\alpha =\sqrt{\frac{T-t}{t(1-\rho ^{2})}}\rho -1,$ $k=\frac{T-t}{%
1-\rho ^{2}}$ and $\lambda _{2}$ is such that the budget constraint is
satisfied. Note that the expected values are proportional to the
probabilities to beat the target value $b$. We observe that in the
constrained target probability maximization problem the expected value (and
the corresponding success probability) is smaller than in the unconstrained
problem. 

\section{Final remarks\label{Final}}

In this paper, we introduce a state-dependent version of the optimal
investment problem. We deal with investors who target a known wealth
distribution at maturity (as in the traditional setting) and \textit{%
additionally} desire a particular interaction with a random benchmark. We
show that optimal contracts depend at most on two underlying assets, or on
one asset evaluated at two different dates, and we are able to characterize
and determine them explicitly. Our characterization of optimal strategies
allows us to extend the classical expected utility optimization problem of
Merton to the state-dependent situation. Throughout the paper, we have
assumed that the state-price density process $\xi _{T}$ is a decreasing
functional of the risky asset price $S_{T}$ and that there is a single risky
asset. It is possible to relax these assumptions and yet still to provide
explicit representations of optimal payoffs. However, the optimality is then
no longer related to path-independence properties.

Throughout the paper, we assumed that $\xi _{T}$ is decreasing in $S_{T}$
(in \eqref{function}). Moreover, we use the one-dimensional Black-Scholes
model to illustrate our findings. However, the case of multidimensional
markets described by a price process $(S_{t}^{(1)},\dots ,S_{t}^{(d)})_{t}$
is essentially included in the results presented in this paper, assuming
that the state-price density process $(\xi _{t})_{t}$ of the risk-neutral
measure chosen for pricing is of the form $\xi _{t}=g_{t}\left(
h_{t}(S_{t}^{(1)},\dots ,S_{t}^{(d)})\right) $ with some real functions $%
g_{t}$, $h_{t}$ (as in Bernard, Maj and Vanduffel \citeyear{BMV} who
considered the state-independent case). All results in the paper apply by
replacing the one-dimensional stock price process $S_{t}$ by the
one-dimensional process $h_{t}(S_{t}^{(1)},\dots ,S_{t}^{(d)})$. In
addition, we have assumed that asset prices are continuously distributed,
which amounts essentially to assuming that the state-price density process $%
\xi _{t}$ is continuously distributed at any time. An extension to the case
in which $\xi _{t}$ may have atoms is possible but not in the scope of the
present paper.

A straightforward extension of the results presented in this paper is to
consider the market model of Platen and Heath \citeyear{PlatenHeath2006}
using the Growth Optimal Portfolio (GOP). Its origins can be traced back to
Kelly \citeyear{K}. It consists of replacing the state-price density process 
$\xi _{t}$ by $1/S_{t}^{\ast }$, where $S_{t}^{\ast }$ denotes the value of
the GOP at time $t$. In the Black-Scholes setting, $S_{t}^{\ast }$ is simply
the value of one unit investment in a constant-mix strategy, where a
fraction $\frac{\theta }{\sigma }$ is invested in the risky asset and the
remaining fraction $1-\frac{\theta }{\sigma }$ in the bank account. 
It is easy to prove that this strategy is optimal for an expected
log-utility maximizer. Using a milder notion of arbitrage, Platen and Heath %
\citeyear{PlatenHeath2006} argue that, in general, the price of
(non-negative) payoffs could be achieved using the pricing rule (\ref%
{pricing}) where the role of $\xi _{T}$ is now played by the inverse of the
GOP. Hence, our results are also valid in their setting, where the GOP is
taken as the reference (see Bernard et al. \citeyear{BCVQF} for an example). 
\newpage \appendix

\section{Proofs}

%
%
%
%
%
%
%
%
Throughout the paper and the different proofs, we make repeatedly use of the
following lemmas. The first lemma gives a restatement of the classical
Hoeffding--Fr\'{e}chet bounds going back to the early work of Hoeffding %
\citeyear{Hoef} and Fr\'{e}chet \citeyear{F-40,F}.

\begin{lem}[Hoeffding--Fr\'{e}chet bounds]
\label{lem1} Let $(X,Y)$ be a random pair and $U$ uniformly distributed on $%
(0,1)$. Then 
\begin{equation}
\mathbb{E}\left[ F_{X}^{-1}(U)F_{Y}^{-1}(1-U)\right] \leq \mathbb{E}\left[ XY%
\right] \leq \mathbb{E}\left[ F_{X}^{-1}(U)F_{Y}^{-1}(U)\right] .
\label{EQ2}
\end{equation}%
The upper bound for $\mathbb{E}\left[ XY\right] $ is attained if and only if 
$(X,Y)$ is comonotonic, i.e. $(X,Y)\sim (F_{X}^{-1}(U),F_{Y}^{-1}(U)).$
Similarly, the lower bound for $\mathbb{E}\left[ XY\right] $ is attained if
and only if $(X,Y)$ is anti-monotonic, i.e. $(X,Y)\sim
(F_{X}^{-1}(U),F_{Y}^{-1}(1-U)).$
\end{lem}

The following lemma combines special cases of two classical construction
results. The Rosenblatt transformation describes a transform of a random
vector to iid uniformly distributed random variables (see Rosenblatt %
\citeyear{R-52}). The second result is a special form of the standard
recursive construction method for a random vector with given distribution
out of iid uniform random variables due to O'Brien \citeyear{O-75}, Arjas
and Lehtonen \citeyear{AL} and R\"uschendorf \citeyear{Rue-81}.

\begin{lem}[Construction method]
\label{lem2} Let $(X,Y)$ be a random pair and assume that $F_{Y|X=x}(\cdot )$
is continuous $\forall x$. Denote $V=F_{Y|X}(Y).$ Then $V$ is uniformly
distributed on $(0,1)$ and independent of $X.$ It is also increasing in $Y$
conditionally on $X$. Furthermore, for every variable $Z$, $(X,$ $%
F_{Z|X}^{-1}(V))\sim (X,Z).$
\end{lem}

For the proof of the first part note that by the continuity assumption on $%
F_{Y\mid X=x}$ we get from the standard transformation 
\begin{equation*}
\left( V\mid X=x\right) \sim \left( F_{Y\mid X=x}(Y)\mid X=x\right) \sim
U(0,1),\quad \forall x.
\end{equation*}%
Clearly $V\sim U(0,1).$ Furthermore, the conditional distribution $F_{V\mid
X=x}$ does not depend on $x$ and thus $V$ and $X$ are independent. For the
second part one gets by the usual quantile construction that $F_{Z\mid
X=x}^{-1}(V)$ has distribution function $F_{Z\mid X=x}$. This implies that $%
(X,F_{Z\mid X}^{-1}(V))$ $\sim $ $(X,Z)$ since both sides have the same
first marginal distribution and the same conditional distribution.

\medskip

\begin{lem}
\label{lem3} Let ($X,Y)$ be jointly normally distributed. Then,
conditionally on $Y,$ $X$ is normally distributed and, 
\begin{eqnarray*}
\mathbb{E}(X|Y) &=&\mathbb{E}(X)+\frac{\mathop{\mathrm{cov}}(X,Y)}{%
\mathop{\mathrm{var}} (Y)}(Y-\mathbb{E}(Y) \\
var(X|Y) &=&(1-\rho ^{2})\mathop{\mathrm{var}} (X).
\end{eqnarray*}%
Denote the density of $Y$ by $f_{Y}(y)$. One has, 
\begin{equation*}
\int_{-\infty }^{c}e^{a+by}f_{Y}(y)dy=e^{a+b\mathbb{E}(Y)+\frac{b^{2}}{2}%
\mathop{\mathrm{var}} (Y)}\frac{1}{\sqrt{2\pi \mathop{\mathrm{var}} (Y)}}%
\int_{-\infty }^{c}e^{-\frac{1}{2}\left( \frac{y-(\mathbb{E}(Y)+b%
\mathop{\mathrm{var}} (Y))}{\sqrt{\mathop{\mathrm{var}} (Y)}}\right) ^{2}}dy.
\end{equation*}
\end{lem}

\noindent The results in this lemma are well-known and we omit its
proof.\hfill $\Box $

\subsection{Proof of Proposition \protect\ref{pr3b}\label{proofpr3b}}

Let $U=F_{S_{T}}(S_{T})$ a uniformly distributed variable on $(0,1).$
Consider a payoff $X_{T}$. One has, 
\begin{eqnarray*}
c_{0}(X_{T}) &=&\mathbb{E}\left[ X_{T}\xi _{T}\right] \geq \mathbb{E}\left[
F_{X_{T}}^{-1}(U)\xi _{T}\right] =c_{0}(X_{T}^{{*} }),
\end{eqnarray*}%
where the inequality follows from the fact that $F_{X_{T}}^{-1}(U)$ and $\xi
_{T}$ are anti-monotonic and using the Hoeffding--Fr\'echet bounds in Lemma %
\ref{lem1}. Hence, $X_{T}^{*}=F^{-1}(F_{S}(S_{T}))$ is the cheapest payoff
with cdf $F.$ Similarly, the most expensive payoff with cdf $F$ writes as $%
Z_{T}^{{*} }=F^{-1}(1-F_{S}(S_{T}))$. Since $c$ is the price of a payoff $%
X_{T}$ with cdf $F$, one has%
\begin{equation*}
c\in \lbrack c_{0}(X_{T}^{{*} }), c_{0}(Z_{T}^{{*} })].
\end{equation*}%
If $c=c_{0}(X_{T}^{{*} })$ then $X_{T}^{{*} }$ is a solution. Similarly, if $%
c=c_{0}(Z_{T}^{{*} })$ then $Z_{T}^{{*} }$ is a solution. Next, let $c\in
(c_{0}(X_{T}^{{*} }),c_{0}(Z_{T}^{{*} }))$ and define the payoff $%
f_{a}(S_{T})$ with $a\in \mathbb{R}$, 
\begin{equation*}
f_{a}(S_{T})=F^{-1}\left[ (1-F_{S_{T}}(S_{T}))\mathds{1}_{S_{T}\leq
a}+(F_{S_{T}}(S_{T})-F_{S_{T}}(a))\mathds{1}_{S_{T}>a}\right] .
\end{equation*}%
Then $f_{a}(S_{T})$ is distributed with cdf $F$. The price $%
c_{0}(f_{a}(S_{T}))$ of this payoff is a continuous function of the
parameter $a$. Since $\lim_{a\rightarrow
0^{+}}c_{0}(f_{a}(S_{T}))=c_{0}(X_{T}^{{*} })$ and $\lim_{a\rightarrow
+\infty }c_{0}(f_{a}(S_{T}))=c_{0}(Z_{T}^{{*} })$, using the theorem of
intermediary values for continuous functions, there exists $a^{%
\raisebox{-.5ex}{\footnotesize *}}$ such that $c_{0}(f_{a^{%
\raisebox{-1.5ex}{\scriptsize *}}}(S_{T}))=c$. This ends the proof.{\hfill} $%
\Box $

\subsection{Proof of Corollary \protect\ref{CEp}\label{proofCEp}}

Let $X_{T}\sim F$ be cost-efficient. Then $X_{T}$ solves (\ref{CONSTEFF0})
and Theorem \ref{theo:pr1} implies that $X_{T}=F^{-1}(F_{S_{T}}(S_{T}))$
almost surely. Reciprocally, let $X_{T}\sim F$ be increasing in $S_{T}. $
Then, by our continuity assumption, $X_{T}=F^{-1}(F_{S_{T}}(S_{T}))$ almost
surely and thus $X_{T}$ is cost-efficient. \hfill $\Box $

\subsection{Proof of Theorem \protect\ref{main}\label{proofmain}}

The idea of the proof is very similar to the proof of Proposition \ref{pr3b}%
. Let $U$ be given by $U=F_{S_{T}|A_{T}}(S_{T}).$ It is uniformly
distributed over $(0,1)$ and independent of $A_{T}$ (see Lemma \ref{lem2}).
Furthermore, conditionally on $A_{T},$ $U$ is increasing in $S_{T}$.
Consider next a payoff $X_{T}$ and note that $F_{X_{T}|A_{T}}^{-1}(U)\sim
X_{T}.$ We find that 
\begin{eqnarray}
c_{0}(X_{T}) &=&\mathbb{E}\left[ X_{T}\xi _{T}\right] =\mathbb{E}[\mathbb{E}%
[\left. X_{T}\xi _{T}\right\vert A_{T}]]  \notag \\
&\geq &\mathbb{E}\left[ \mathbb{E}\left[ \left. F_{X_{T}|A_{T}}^{-1}(U)\xi
_{T}\right\vert A_{T}\right] \right] =\mathbb{E}\left[
F_{X_{T}|A_{T}}^{-1}(U)\xi _{T}\right],  \label{mini}
\end{eqnarray}%
where the inequality follows from the fact that $F_{X_{T}|A_{T}}^{-1}(U)$
and $\xi _{T}$ are conditionally (on $A_{T}$) anti-monotonic and using %
\eqref{EQ2} in Lemma \ref{lem1} for the conditional expectation
(conditionally on $A_{T}$). Similarly, one finds that 
\begin{equation*}
c_{0}(X_{T})\leq \mathbb{E}\left[ F_{X_{T}|A_{T}}^{-1}(1-U)\,\xi _{T}\right].
\end{equation*}%
Next we define the uniform $(0,1)$ distributed variable,%
\begin{equation*}
g_{a}(S_{T})=(1-F_{S_{T}}(S_{T}))\mathds{1}_{S_{T}\leq
a}+(F_{S_{T}}(S_{T})-F_{S_{T}}(a))\mathds{1}_{S_{T}>a}.
\end{equation*}%
We observe that thanks to Lemma \ref{lem2}, $%
F_{g_{a}(S_{T})|A_{T}}(g_{a}(S_{T}))$ is independent of $A_{T}$ and also
that $f_{a}\left( S_{T},A_{T}\right) $ given as%
\begin{equation*}
f_{a}\left( S_{T},A_{T}\right)
=F_{X_{T}|A_{T}}^{-1}(F_{g_{a}(S_{T})|A_{T}}(g_{a}(S_{T})))
\end{equation*}%
is a twin with the desired joint distribution $G$ with $A_{T}.$ Denote by $%
X_{T}^{*} =F_{X_{T}|A_{T}}^{-1}(U)$ and by $Z_{T}^{{*}}
=F_{X_{T}|A_{T}}^{-1}(1-U)$. Note that $X_{T}^{*}
=f_{0}\left(S_{T},A_{T}\right) $ and $Z_{T}^{*}=f_{1}\left(
S_{T},A_{T}\right) $ almost surely. The same discussion as in the proof of
Proposition \ref{pr3b} applies here. When $c=c_{0}(X_{T}^{*})$ then $%
X_{T}^{*}$ is a twin with the desired properties. Similarly, when $%
c=c_{0}(Z_{T}^{*})$ then $Z_{T}^{*}$ is a twin with the desired properties.
Otherwise, when $c\in (c_{0}(X_{T}^{*}),c_{0}(Z_{T}^{*}))$ then the
continuity of $c_{0}(f_{a}\left( S_{T},A_{T}\right) )$ with respect to $a$
ensures that there exists $a^{*}$ such that $c:=c_{0}(f_{a^{*}}\left(
S_{T},A_{T}\right) )$. Thus, $f_{a^{*}}\left( S_{T},A_{T}\right) $ is a twin
with the desired joint distribution $G$ with $A_{T}$ and with cost $c$. This
ends the proof. \hfill $\Box $

\subsection{Proof of Theorem \protect\ref{twin2}\label{prooftwin2}}

Let $0<t<T.$ It follows from Lemma \ref{lem2} that $F_{S_{t}|S_{T}}(S_{t})$
is uniformly distributed on $(0,1)$ and independent of $S_{T}.$ Let the twin 
$f(S_{t},S_{T})$ be given as 
\begin{equation*}
f(S_{t},S_{T}):=F_{X_{T}|S_{T}}^{-1}(F_{S_{t}|S_{T}}(S_{t})).
\end{equation*}%
Using Lemma \ref{lem2} again, one finds that $(f(S_{t},S_{T}),S_{T})\sim
(X_{T},S_{T})\sim G.$ This also implies, 
\begin{eqnarray*}
c_{0}(f(S_{t},S_{T})) &=&\mathbb{E}\left[ f(S_{t},S_{T})\xi _{T}\right]=%
\mathbb{E[}X_{T}\xi _{T}]=c_{0}(X_{T}),
\end{eqnarray*}%
and this ends the proof.\hfill $\Box $

\subsection{Proof of Theorem \protect\ref{main2}\label{proofmain2}}

It follows from Lemma \ref{lem2} that $U=F_{S_{T}|A_{T}}(S_{T})$ is
uniformly distributed on $(0,1)$, stochastically independent of $A_{T}$ and
increasing in $S_{T}$ conditionally on $A_T$. Let the twin $X_{T}^{*}$ be
given as 
\begin{equation*}
X_{T}^{*}=F_{X_{T}|A_{T}}^{-1}(U).
\end{equation*}%
Invoking Lemma \ref{lem2} again, $(X_{T}^{*},A_{T})\sim (X_{T},A_{T})\sim G.$
Moreover, 
\begin{eqnarray*}
c_{0}(X_{T}) =\mathbb{E}\left[ X_{T}\xi _{T}\right] &=&\mathbb{E}[\mathbb{E}%
[\left. X_{T}\xi _{T}\right\vert A_{T}]] \\
&\geq &\mathbb{E}\left[ \mathbb{E}\left[ \left. F_{X_{T}|A_{T}}^{-1}(U)\xi
_{T}\right\vert A_{T}\right] \right] \\
&=&\mathbb{E}\left[ F_{X_{T}|A_{T}}^{-1}(U)\xi _{T}\right] =c_{0}(X_{T}^{*})
\end{eqnarray*}%
where the inequality follows from the fact that $F_{X_{T}|A_{T}}^{-1}(U)$
and $S_{T}$ are conditionally (on $A_{T}$) comonotonic and using \eqref{EQ2}
in Lemma \ref{lem1} for the conditional expectation (conditionally on $A_{T}$%
). \hfill $\Box $

\subsection{Proof of Corollary \protect\ref{coro2} \label{proofcoro2}}

Let us first assume that $X_{T}$ is a cheapest twin. By Theorem \ref{main2}, 
$X_{T}$ is (almost surely)\ equal to $X_{T}^{{*} }$ as defined by %
\eqref{xtstar}\ which is, conditionally on $A_{T}$, increasing in $S_{T}$.
Reciprocally, we now assume that $X_{T}=f\left( S_{T},A_{T}\right) $ is
conditionally on $A_{T}$ increasing in $S_{T}$. Hence $%
X_{T}=F_{X_{T}|A_{T}}^{-1}\left( F_{S_{T}|A_{T}}\left( S_{T}\right) \right) $
almost surely, which means it is a solution to (\ref{CONSTEFF}) and thus a
cheapest twin%
%
%
. \hfill $\Box $

\section{Security design}

\subsection{Twin of the fixed strike (continuously monitored)\ geometric
Asian call option\label{Geom}}

Expression \eqref{twinprop32} allows us to find twins satisfying the
constraint \eqref{JOINT} on the dependence with the benchmark $S_{T}$. Using
Lemma \ref{lem3} we find that%
\begin{equation*}
\ln (S_{t}/S_0)|\ln (S_{T}/S_0)\sim \mathcal{N}\left( \frac{t}{T}\ln \left( 
\frac{S_{T}}{S_{0}}\right),\sigma ^{2}t\left( 1-\frac{t}{T}\right) \right),
\end{equation*}%
and thus%
\begin{equation*}
{F_{S_{t}|S_{T}}(S_{t})}=\Phi \left( \frac{\ln \left( \frac{S_{t}S_{0}^{%
\frac{t}{T}-1}}{S_{T}^{\frac{t}{T}}}\right) }{\sigma \sqrt{\frac{tT-t^{2}}{T}%
}}\right) .
\end{equation*}%
Furthermore, the couple ($\ln \left( G_{T}\right),\ln \left( S_{T}\right) )$
is bivariate normally distributed with mean and variance for the marginal
distributions that are given as $\mathbb{E}[\ln (G_{T})]=\ln S_{0}+\left(
\mu -\frac{1}{2} \sigma ^{2}\right) \frac{T}{2}$, $\hbox{var}\lbrack \ln
(G_{T})]=\frac{\sigma ^{2}T}{3}$ and $\mathbb{E}[\ln (S_{T})]=\ln
S_{0}+\left( \mu -\frac{1}{2}\sigma ^{2}\right) {T}$, $\hbox{var}\lbrack \ln
(S_{T})]=\ {\sigma ^{2}T} $. For the correlation coefficient one has $\rho
(\ln (S_{T}),\ln (G_{T}))= \frac{\sqrt{3}}{2}.$ Applying Lemma \ref{lem3}
again one finds that,%
\begin{equation}
\ln (G_{T})|\ln (S_{T})\sim \mathcal{N}\left(\ln \left(
S_{0}^{1/2}S_{T}^{1/2}\right),\frac{\sigma ^{2}T}{12}\right),  \label{G|S}
\end{equation}%
and thus,%
\begin{equation*}
F_{G_{T}|S_{T}}(x)=\Phi \left( \frac{\ln (x)-\ln \left(
S_{0}^{1/2}S_{T}^{1/2}\right) }{\frac{\sigma \sqrt{T}}{2\sqrt{3}}}\right) .
\end{equation*}%
Therefore, 
\begin{equation*}
F_{G_{T}|S_{T}}^{-1}(y)=\exp \left( \ln \left( S_{0}^{1/2}S_{T}^{1/2}\right)
+\frac{\sigma \sqrt{T}}{2\sqrt{3}}\Phi ^{-1}(y)\right) .
\end{equation*}%
The expression of $R_{T}(t)$ given in \eqref{RT} is then straightforward to
derive.

\hspace*{-1.5ex}For choosing a specific twin among others, we suggest to
maximize $\rho \left( \ln R_{T} ( t),\ln G_{T}\right)$. First, we calculate,%
\begin{eqnarray*}
\mathop{\mathrm{cov}}\left( \ln S_{T},\frac{1}{T}\int_{0}^{T}\ln
\left(S_{s}\right) ds\right) &=&\frac{1}{T}\int_{0}^{T}\mathop{\mathrm{cov}}%
\left( \ln S_{T}, \ln \left( S_{s}\right) \right) ds \\
&=&\frac{\sigma ^{2}}{T} \int_{0}^{T}\left( s\wedge T\right) ds=\frac{\sigma
^{2}T}{2}.
\end{eqnarray*}%
%
%
%
%
%
%
%
%
%
%
%
%
%
%
%
%
%
%
%
%
%
%
%
%
%
%
%
%
%
%
%
%
%
%
%
%
%
%
%
%
%
%
%
%
%
%
%
%
%
%
%
%
%
%
%
%
%
%
%
%
%
%
%
%
%
%
%
%
%
%
%
%
Furthermore, by denoting $a=\frac{1}{2}-\frac{1}{2\sqrt{3}}\sqrt{\frac{T-t}{t%
}}$, $b=\frac{T}{t}\frac{1}{2\sqrt{3}}\sqrt{\frac{t}{T-t}}$ and $c=\frac{1}{2%
}-\frac{1}{2\sqrt{3}}\sqrt{\frac{t}{T-t}}$, equation \eqref{RT} may be
rewritten as $\ln R_{T}\left( t\right) =a\ln S_{0}+b\ln S_{t}+c\ln S_{T}$.
The covariance being bilinear, one then has, 
\begin{eqnarray*}
{\mathop{\mathrm{cov}}\left( \ln R_{T}\left( t\right), \ln G_{T}\right) }&=&
b\mathop{\mathrm{cov}}\left( \ln S_{t}, \frac{1}{T}\int_{0}^{T}\ln \left(
S_{s}\right) ds\right) +c\mathop{\mathrm{cov}}\left( \ln S_{T}, \frac{1}{T}%
\int_{0}^{T}\ln \left( S_{s}\right) ds\right) \\
&=&\frac{\sigma ^{2}}{2}\left( \frac{T}{2}+\frac{\sqrt{t}\sqrt{T-t}}{2\sqrt{3%
}}\right) .
\end{eqnarray*}%
Denote by $\sigma _{\ln R_{T}\left( t\right) }$ and by $\sigma _{\ln G_T }$
the respective standard deviations. For the correlation we find that 
\begin{eqnarray*}
\rho \left( \ln R_{T}\left( t\right), \ln G_{T}\right) =\frac{%
\mathop{\mathrm{cov}}\left( \ln R_{T}\left( t\right), \ln G_{T}\right) }{%
\sigma _{\ln R_{T}\left( t\right) }\sigma _{\ln G_{T}}} &=&\frac{3}{4}+\frac{%
\sqrt{3}\sqrt{\left( T-t\right) t}}{4T}.
\end{eqnarray*}%
%
%
%
%
%
%
%
%
%
%
%
%
%
%
%
%
%
%
%
%
%
%
%
%
%
%
%
%
%
%
%
%
%
%
%
%
%
%
%
%
%
%
%
%
%
%
%
%
%
%
%
%
%
%
%
%
%
%
%
%
%
%
%
%
%
%
%
%
%
%
%
%
%
%
%
%
%
%
%
%
%
%
Hence $\rho \left( \ln R_{T}\left( t\right), \ln G_{T}\right) $ is maximized
for $t=\frac{T}{2}$. \hfill $\Box $

\subsection{Twin of the floating strike (continuously monitored) geometric
Asian put option\label{Geom2}}

We first recall from equation (\ref{G|S}) that, 
\begin{equation*}
\ln (G_{T})|\ln (S_{T})\sim \mathcal{N}\left( \ln \left( S_{0}^{\frac{1}{2}%
}S_{T}^{\frac{1}{2}}\right),\frac{\sigma ^{2}T}{12}\right) .
\end{equation*}%
Therefore $Y_{T}=(G_{T}-S_{T})^{+}$ has the following conditional cdf 
\begin{equation*}
\P (Y_{T}\leq y|S_{T}=s)= \Phi \left( \frac{\ln (s+y)-\ln \left(
S_{0}^{1/2}s^{1/2}\right) }{\frac{\sigma \sqrt{T}}{2\sqrt{3}}}\right)%
\mathds{1}_{y\geq 0}
\end{equation*}%
Then 
\begin{equation*}
F_{Y_{T}|S_{T}}^{-1}(z)=\left( S_{0}^{\frac{1}{2}}S_{T}^{\frac{1}{2}}e^{%
\frac{\sigma }{2}\sqrt{\frac{T}{3}}\Phi ^{-1}(z)}-S_{T}\right) ^{+}.
\end{equation*}%
Therefore $F_{Y_{T}|S_{T}}^{-1}\left( F_{S_{t}|S_{T}}(S_{t}))\right)$ can
then easily be computed and after some calculations it simplifies to %
\eqref{twi}.\hfill $\Box $

\subsection{Cheapest Twin of the floating strike (continuously monitored)
geometric Asian put option\label{Geom3}}

%
%
%
%
%
Applying Lemma \ref{lem3} we find,%
\begin{equation*}
\ln (S_{T})|\ln (G_{T})\sim \mathcal{N}\left( \ln \left( \frac{G_{T}^{3/2}}{%
S_{0}^{\frac{1}{2}}}\right) +\frac{1}{4}\left( \mu -\frac{\sigma ^{2}}{2}%
\right) T,\ \frac{\sigma ^{2}T}{4}\right) .
\end{equation*}%
Hence, 
\begin{equation}
{F_{S_{T}|G_{T}}(S_{T})})=\Phi \left( \frac{\ln \left( \frac{S_{T}S_{0}^{%
\frac{1}{2}}}{G_{T}^{\frac{3}{2}}}\right) -\left( \mu -\frac{\sigma ^{2}}{2}%
\right) \frac{T}{4}}{\frac{\sigma \sqrt{T}}{2}}\right) .  \label{stst}
\end{equation}%
Furthermore, $Y_{T}=(G_{T}-S_{T})^{+}$ has the following conditional cdf,%
\begin{equation*}
P(Y_{T}\leq y|G_{T}=g)=\left\{ 
\begin{array}{cl}
1 & \hbox{if}\ y\geq g, \\ 
\Phi \left( \frac{\ln \left( \frac{g^{3/2}}{S_{0}^{\frac{1}{2}}}\right) +%
\frac{1}{4}\left( \mu -\frac{\sigma ^{2}}{2}\right) T-\ln (g-y)}{\frac{%
\sigma \sqrt{T}}{2}}\right) & \hbox{if}\ 0\leq y\leq g, \\ 
0 & \hbox{if}\ y<0.%
\end{array}%
\right.
\end{equation*}%
Then 
\begin{equation*}
F_{Y_{T}|G_{T}}^{-1}(z)=\left( G_{T}-\frac{G_{T}^{\frac{3}{2}}}{S_{0}^{\frac{%
1}{2}}}e^{\frac{1}{4}\left( \mu -\frac{\sigma ^{2}}{2}\right) T-\frac{\sigma 
}{2}\sqrt{T}\Phi ^{-1}(z)}\right) ^{+}.
\end{equation*}%
Replacing $z$ by the expression \eqref{stst} for ${F_{S_{T}|G_{T}}(S_{T})})$
derived above, then gives rise to expression \eqref{OPTI}.\hfill $\Box $

\subsection{Derivation of prices (\protect\ref{float}) and (\protect\ref%
{pricy}) \label{proofpricy}}

\paragraph{Price (\protect\ref{float})}

~\newline
\noindent Let us observe that,%
\begin{equation*}
\begin{array}{rcl}
\left( G_{T}-S_{T}\right) ^{+} & = & G_{T}\left( 1-\frac{S_{T}}{G_{T}}%
\right) ^{+}=S_{0}e^{Y}\left( 1-e^{Z}\right) ^{+},%
\end{array}%
\end{equation*}%
where $Z=X-Y,$ $Y=\ln \left( \frac{G_{T}}{S_{0}}\right) ,$ $X=\ln \left( 
\frac{S_{T}}{S_{0}}\right) $. We find, with respect to the \textit{risk
neutral measure} $\mathbb{Q},$%
\begin{eqnarray*}
\mathbb{E}_{\mathbb{Q}}\left[ \left( G_{T}-S_{T}\right) ^{+}\right] &=&S_{0}%
\mathbb{E}_{\mathbb{Q}}\left( \mathbb{E}_{\mathbb{Q}}\left[ e^{Y}|Z\right]
\left( 1-e^{Z}\right) ^{+}\right) \\
&=&S_{0}\mathbb{E}_{\mathbb{Q}}\left[ \left( e^{\mathbb{E}_{\mathbb{Q}}(Y|Z)+%
\frac{1}{2}\mathop{\mathrm{var}_{\mathbb{Q}}}(Y|Z)}-e^{\mathbb{E}_{\mathbb{Q}%
}(Y|Z)+\frac{1}{2}\mathop{\mathrm{var}_{\mathbb{Q}}}(Y|Z)+Z}\right) ^{+}%
\right] .
\end{eqnarray*}%
We now compute (still with respect to $\mathbb{Q})$,%
\begin{eqnarray*}
\mathbb{E}_{\mathbb{Q}}(Y|Z) &=&\mathbb{E}_{\mathbb{Q}}(Y)+\frac{%
\mathop{\mathrm{cov}_{\mathbb{Q}}}(Y,Z)}{\mathop{\mathrm{var}_{\mathbb{Q}}}%
(Z)}(Z-\mathbb{E}_{\mathbb{Q}}(Z))=\left( r-\frac{\sigma ^{2}}{2}\right) 
\frac{T}{4}+\frac{1}{2}Z \\
\mathop{\mathrm{var}_{\mathbb{Q}}}(Y|Z) &=&(1-\rho ^{2})\mathop{%
\mathrm{var}_{\mathbb{Q}}}(Y)=\frac{3}{4}\frac{\sigma ^{2}T}{3}=\frac{\sigma
^{2}T}{4}.
\end{eqnarray*}%
Hence, 
\begin{eqnarray*}
\mathbb{E}_{\mathbb{Q}}\left( G_{T}-S_{T}\right) ^{+} &=&S_{0}\mathbb{E}_{%
\mathbb{Q}}\left( e^{r\frac{T}{4}+\frac{1}{2}Z}-e^{r\frac{T}{4}+\frac{3}{2}%
Z}\right) ^{+} \\
&=&S_{0}\int_{-\infty }^{0}e^{^{r\frac{T}{4}+\frac{1}{2}Z}}f_{Z}(z)dz-S_{0}%
\int_{-\infty }^{0}e^{r\frac{T}{4}+\frac{3}{2}Z}f_{Z}(z)dz,
\end{eqnarray*}%
where $f_{Z}(z)$ is now denoting the density of $Z$ under $\mathbb{Q}.$ Here 
$Z$ is normally distributed with parameters $(r-\frac{\sigma ^{2}}{2})\frac{T%
}{2}$ and variance $\frac{\sigma ^{2}T}{3}.$ Hence, taking into account
Lemma \ref{lem3},%
\begin{eqnarray*}
{\mathbb{E}_{\mathbb{Q}}\left( G_{T}-S_{T}\right) ^{+}} &=&S_{0}e^{r\frac{T}{%
2}-\sigma ^{2}\frac{T}{12}}\Phi \left( \frac{-\left( r-\frac{\sigma ^{2}}{2}%
\right) \frac{T}{2}-\frac{\sigma ^{2}T}{6}}{\sqrt{\frac{\sigma ^{2}T}{3}}}%
\right) -S_{0}e^{rT}\Phi \left( \frac{-\left( r-\frac{\sigma ^{2}}{2}\right) 
\frac{T}{2}-\frac{\sigma ^{2}T}{2}}{\sqrt{\frac{\sigma ^{2}T}{3}}}\right)
\end{eqnarray*}%
Choose $f=\frac{-r\frac{T}{2}+\frac{\sigma ^{2}T}{12}}{\sigma \sqrt{\frac{T}{%
3}}}$ to obtain \eqref{float}.

\paragraph{Price (\protect\ref{pricy})}

~\newline
\noindent One has,%
\begin{equation*}
\left( G_{T}-a\frac{G_{T}^{3}}{S_{T}}\right) ^{+}=G_{T}\left( 1-a\frac{%
G_{T}^{2}}{S_{T}}\right) ^{+}=S_{0}e^{Y}\left( 1-ce^{Z}\right) ^{+}
\end{equation*}%
where $Z=2Y-X,$ $Y=\ln \left( \frac{G_{T}}{S_{0}}\right) ,$ $X=\ln \left( 
\frac{S_{T}}{S_{0}}\right) ,$ $c=e^{\left( \mu -\frac{\sigma ^{2}}{2}\right) 
\frac{T}{2}}.$ Hence, with respect to the \textit{risk neutral measure} $%
\mathbb{Q},$ 
\begin{eqnarray*}
\mathbb{E}_{\mathbb{Q}}\left( G_{T}-a\frac{G_{T}^{3}}{S_{T}}\right) ^{+}
&=&S_{0}\mathbb{E}_{\mathbb{Q}}\left( \mathbb{E}_{\mathbb{Q}}(e^{Y}|Z)\left(
1-ce^{Z}\right) ^{+}\right) \\
&=&S_{0}\mathbb{E}_{\mathbb{Q}}\left( e^{\mathbb{E}_{\mathbb{Q}}(Y|Z)+\frac{1%
}{2}\mathop{\mathrm{var}_{\mathbb{Q}}}(Y|Z)}-ce^{\mathbb{E}_{\mathbb{Q}%
}(Y|Z)+\frac{1}{2}\mathop{\mathrm{var}_{\mathbb{Q}}}(Y|Z)+Z}\right) ^{+}.
\end{eqnarray*}%
We now compute, 
\begin{equation*}
\mathbb{E}_{\mathbb{Q}}(Y|Z)=\left( r-\frac{\sigma ^{2}}{2}\right) \frac{T}{2%
}+\frac{1}{2}Z\quad \mathrm{and}\quad \mathop{\mathrm{var}_{\mathbb{Q}}}%
(Y|Z)=\frac{\sigma ^{2}T}{4}.
\end{equation*}%
Hence, 
\begin{eqnarray*}
\mathbb{E}_{\mathbb{Q}}\left( G_{T}-a\frac{G_{T}^{3}}{S_{T}}\right) ^{+}
&=&S_{0}\mathbb{E}_{\mathbb{Q}}\left( e^{r\frac{T}{2}-\frac{\sigma ^{2}T}{8}+%
\frac{1}{2}Z}-ce^{r\frac{T}{2}-\frac{\sigma ^{2}T}{8}+\frac{3}{2}Z}\right)
^{+} \\
&=&S_{0}\int_{-\infty }^{\ln (c)}e^{^{r\frac{T}{2}-\frac{\sigma ^{2}T}{8}+%
\frac{1}{2}Z}}f_{Z}(z)dz-S_{0}c\int_{-\infty }^{\ln (c)}e^{r\frac{T}{2}-%
\frac{\sigma ^{2}T}{8}+\frac{3}{2}Z}f_{Z}(z)dz,
\end{eqnarray*}%
where $f_{Z}(z)$ is the density of $Z,$ under $\mathbb{Q}.$ Note that $Z$ is
normally distributed with parameters $0$ and variance $\frac{\sigma ^{2}T}{3}%
.$ Taking into account Lemma \ref{lem3}, 
\begin{eqnarray*}
\mathbb{E}_{\mathbb{Q}}\left( G_{T}-a\frac{G_{T}^{3}}{S_{T}}\right) ^{+} &=&
S_{0}e^{\frac{rT}{2}}\left( \Phi \left( d\right) e^{-\frac{\sigma ^{2}T}{12}%
}-e^{\frac{\mu T}{2}}\Phi \left( d-\frac{\sigma \sqrt{T}}{\sqrt{3}}\right)
\right)
\end{eqnarray*}%
where $d=\frac{-\ln (c)-\frac{\sigma ^{2}T}{6}}{\sigma \sqrt{\frac{T}{3}}}=%
\frac{\frac{\sigma ^{2}T}{12}-\mu \frac{T}{2}}{\sigma \sqrt{\frac{T}{3}}}$%
.\hfill $\Box $

\section{Portfolio Management}

\subsection{Proof of Theorem \protect\ref{EUT2}\label{proofEUT2}}

Let $H_T=\mathbb{E}(\xi_T|Z_T) = \varphi(Z_T)$ and let $\hat\varphi$ denote
the projection of $\varphi$ on the cone $M_\downarrow$ defined as in %
\eqref{Md} with respect to $L^2(\lambda_{[0,1]})$. Then we define $\hat X_T$
and $k(\cdot)$ by 
\begin{eqnarray*}
u^\prime(\hat X_T) := \lambda\hat\varphi(Z_T),
\end{eqnarray*}
i.e. $\hat X_T=\left(u^\prime \right)^{-1}(\lambda\hat\varphi(Z_T))=:k(Z_T)$
with $\lambda$ such that $\mathbb{E}[\xi_T \hat X_T]=\mathbb{E}\left[%
\varphi(Z_T)k(Z_T)\right] = \int_0^1 \varphi(t) k(t)dt=\varphi\cdot k=W_0$.
By definition, $\hat X_T$ is increasing in $Z_T$ since $\left(u^\prime
\right)^{-1}$ is decreasing and $\hat\varphi$ is decreasing (it belongs to $%
M_{\downarrow}$). As a consequence $\hat X_T$ is increasing in $S_T$,
conditionally on $A_T$. For any $Y_T=h(Z_T)$ with a increasing function $h$,
we have by concavity of $u$%
\begin{eqnarray*}
u(Y_T)-u(\hat X_t) &\le& u^\prime (\hat X_T) (Y_T-\hat X_T) = \lambda
\hat\varphi(Z_T)(h(Z_T)-k(Z_T)).
\end{eqnarray*}
Thus, we obtain 
\begin{equation}  \label{eq:C1-1}
\mathbb{E} [u(Y_T)]-\mathbb{E}[u(\hat X_T)] \le \lambda \int_0^1
\hat\varphi(t) (h(t)-k(t))dt = \lambda \hat\varphi\cdot
(h-\Psi(\hat\varphi)),
\end{equation}
where $\Psi(\hat\varphi) =\left(u^\prime\right) ^{-1}(\lambda\hat\varphi) =k$
is increasing and $\Psi(t)=\left(u^\prime\right) ^{-1}(\lambda t)$ is
decreasing.

Now we use some properties of isotonic approximations (see Barlow \textit{et
al.} \citeyear{BBBB}) and obtain 
\begin{eqnarray*}
\hat{\varphi}\cdot (h-\Psi (\hat{\varphi})) &=&\hat{\varphi}\cdot ((-\Psi )(%
\hat{\varphi})-(-h)) \\
&=&\varphi \cdot (-\Psi )(\hat{\varphi})-\hat{\varphi}\cdot (-h)\qquad \text{%
{\small (see Theorem 1.7 in Barlow \textit{et al.} \citeyear{BBBB})}} \\
&=&\varphi \cdot (-h)-\hat{\varphi}\cdot (-h)\quad \phantom{x}\text{{\small %
both claims have price $W_{0}$}} \\
&=&(\varphi -\hat{\varphi})\cdot (-h)\leq 0
\end{eqnarray*}%
by the projection equation (see Theorem 7.8 in Barlow \textit{et al.} %
\citeyear{BBBB}) using that $-h\in M_{\downarrow }$. As a result we obtain
from \eqref{eq:C1-1} that 
\begin{equation*}
\mathbb{E}[u(Y_{T})]\leq \mathbb{E}[u(\hat{X}_{T})],
\end{equation*}%
i.e. $\hat{X}_{T}$ is an optimal claim. \hfill $\Box $

\subsection{Proofs of equations (\protect\ref{XStarCRRA}) and (\protect\ref%
{XHatCRRA}) in the example of subsection \protect\ref{PM EUM}\label%
{ProofExample1}}

We apply Theorem \ref{EUT1} to an investor with a power-utility. Then, 
\begin{equation}
X_{T}^{\star }\left( \eta \right) =(u^{\prime })^{-1}(\lambda \xi
_{T})=(\lambda \xi _{T})^{-\frac{1}{\eta }}  \label{CRRA1}
\end{equation}%
where $\lambda $ is chosen to meet the budget constraint, i.e. 
\begin{equation}
\mathbb{E}[\xi _{T}(u^{\prime })^{-1}(\lambda \xi _{T})]=\mathbb{E}\left[
\xi _{T}(\lambda \xi _{T})^{-\frac{1}{\eta }}\right] =\lambda ^{-\frac{1}{%
\eta }}\mathbb{E}\left[ \xi _{T}^{1-\frac{1}{\eta }}\right] =W_{0}
\label{CRRA2}
\end{equation}%
Since $\xi _{T}=\exp \left\{ -rT-\frac{1}{2}\theta ^{2}T-\theta
Z_{T}\right\} $, 
we find that $\lambda ^{-\frac{1}{\eta }}=W_{0}\exp \left\{ -r\left( \frac{%
1-\eta }{\eta }\right) T-\frac{1}{2}\theta ^{2}T\left( \frac{1-\eta }{\eta }%
\right) \frac{1}{\eta }\right\} $ and 
\begin{equation*}
X_{T}^{\star }\left( \eta \right) =(\lambda \xi _{T})^{-\frac{1}{\eta }%
}=W_{0}e^{-r\left( \frac{1-\eta }{\eta }\right) T-\frac{1}{2}\theta
^{2}T\left( \frac{1-\eta }{\eta }\right) \frac{1}{\eta }-\frac{1}{\eta }%
\left[ \frac{\theta }{\sigma }\left( \mu -\frac{\sigma ^{2}}{2}\right)
T-\left( r+\frac{\theta ^{2}}{2}\right) T\right] }\left( \frac{S_{T}}{S_{0}}%
\right) ^{\frac{\theta }{\sigma \eta }},
\end{equation*}%
which can be simplified to find (\ref{XStarCRRA}). 

Next, we apply Theorem \ref{EUT2} with $A_{T}=S_{t}$, for some $t$ such that 
$t<T$. From Lemma \ref{lem3} we know%
\begin{equation*}
\ln (S_{T})|\ln (S_{t})\sim \mathcal{N}\left( \ln \left( St\right) +\left(
\mu -\frac{\sigma ^{2}}{2}\right) (T-t),\sigma ^{2}(T-t)\right)
\end{equation*}%
so that%
\begin{equation*}
{F_{S_{T}|S_{t}}(S_{T})}=\Phi \left( \frac{\ln \left( \frac{S_{T}}{S_{t}}%
\right) -\left( \mu -\frac{\sigma ^{2}}{2}\right) (T-t)}{\sigma \sqrt{T-t}}%
\right) .
\end{equation*}%
Because $C$ is a Gaussian copula, one has%
\begin{equation*}
C_{1|S_{t}}(x)=\Phi \left[ \frac{\Phi ^{-1}\left[ x\right] -\rho \left( 
\frac{\ln \left( \frac{S_{t}}{S_{0}}\right) -\left( \mu -\frac{\sigma ^{2}}{2%
}\right) t}{\sigma \sqrt{t}}\right) }{\sqrt{1-\rho ^{2}}}\right]
\end{equation*}%
and 
\begin{equation*}
C_{1|S_{t}}^{-1}(y)=\Phi \left[ \sqrt{1-\rho ^{2}}\Phi ^{-1}\left[ y\right]
+\rho \left( \frac{\ln \left( \frac{S_{t}}{S_{0}}\right) -\left( \mu -\frac{%
\sigma ^{2}}{2}\right) t}{\sigma \sqrt{t}}\right) \right] .
\end{equation*}%
This implies 
\begin{equation*}
\zeta _{T}=C_{1|S_{t}}^{-1}({F_{S_{T}|S_{t}}(S_{T})})=\Phi \left[ \varpi _{T}%
\right] ,
\end{equation*}%
where\ $\varpi _{T}$ is a function of $S_{T}$ and $S_{t}$ given by 
\begin{equation}
\varpi _{T}=\sqrt{1-\rho ^{2}}\left( \frac{\ln \left( \frac{S_{T}}{S_{t}}%
\right) -\left( \mu -\frac{\sigma ^{2}}{2}\right) (T-t)}{\sigma \sqrt{T-t}}%
\right) +\rho \left( \frac{\ln \left( \frac{S_{t}}{S_{0}}\right) -\left( \mu
-\frac{\sigma ^{2}}{2}\right) t}{\sigma \sqrt{t}}\right) .  \label{WT}
\end{equation}%
Since $\xi _{T}=\alpha _{T}\left( \frac{S_{T}}{S_{0}}\right) ^{-\beta }$
where $\alpha _{T}=\exp \left( \frac{\theta }{\sigma }\left( \mu -\frac{%
\sigma ^{2}}{2}\right) T-\left( r+\frac{\theta ^{2}}{2}\right) T\right) $, $%
\beta =\frac{\theta }{\sigma }$ and $\theta =\frac{\mu -r}{\sigma }$ (from %
\eqref{xiTform}), one has%
\begin{equation*}
H_{T}=\mathbb{E}(\xi _{T}|\zeta _{T})=\mathbb{E}(\xi _{T}|\varpi
_{T})=\delta e^{-\beta \mathop{\mathrm{cov}}\left( \ln \left( S_{T}\right)
,\varpi _{T}\right) \varpi _{T}},
\end{equation*}%
for some $\delta >0$ and we find%
\begin{equation*}
H_{T}=\delta e^{-\theta \left( \rho \sqrt{t}+\sqrt{(1-\rho ^{2})(T-t)}%
\right) \varpi _{T}}.
\end{equation*}%
Note that conditions on the correlation coefficient imply that $H_{T}$ is
decreasing in $\varpi _{T}$ and thus $H_{T}$ is decreasing in $Z_{T}.$ The
optimal contract thus writes as 
\begin{equation}
\hat{X}_{T}:=\left( u^{\prime }\right) ^{-1}\left( \lambda e^{-\theta \left(
\rho \sqrt{t}+\sqrt{(1-\rho ^{2})(T-t)}\right) \varpi _{T}}\right) ,
\label{XHat}
\end{equation}%
where $\lambda $ is chosen to meet the budget constraint.

When the investor has a power-utility, i.e., $u\left( x\right) =\frac{%
x^{1-\eta }}{1-\eta }$ so that $(u^{\prime })^{-1}\left( x\right) =x^{-\frac{%
1}{\eta }}$ we find that equation (\ref{XHat}) reads as 
\begin{equation}
\hat{X}_{T}\left( \eta \right) :=\lambda ^{-\frac{1}{\eta }}e^{\frac{1}{\eta 
}\theta \left( \rho \sqrt{t}+\sqrt{(1-\rho ^{2})(T-t)}\right) \varpi _{T}}
\label{XHatPowerUtilitywithLambda}
\end{equation}%
and the budget constraint (i.e., $\mathbb{E}\left[ \xi _{T}\hat{X}_{T}\left(
\eta \right) \right] =W_{0}$) requires that 
\begin{equation*}
\mathbb{E}\left[ e^{-rT}e^{-\frac{\theta ^{2}}{2}T-\theta Z_{T}}\lambda ^{-%
\frac{1}{\eta }}\exp \left[ \frac{1}{\eta }\theta \left( \rho \sqrt{t}+\sqrt{%
(1-\rho ^{2})(T-t)}\right) \varpi _{T}\right] \right] =W_{0},
\end{equation*}%
where we have used the expressions for $\xi _{T}$ and $\hat{X}_{T}\left(
\eta \right) $. 
We find that 
\begin{equation*}
\lambda ^{-\frac{1}{\eta }}=W_{0}e^{rT}e^{\theta ^{2}\left( \rho \sqrt{t}+%
\sqrt{(1-\rho ^{2})(T-t)}\right) ^{2}\left( \frac{1}{\eta }-\frac{1}{2\eta
^{2}}\right) }.
\end{equation*}%
The optimal solution is then derived by using this expression into (\ref%
{XHatPowerUtilitywithLambda}). \hfill $\Box $

\subsection{Proof of Proposition \protect\ref{BrP}\label{proofBr}}

Assume that there exists an optimal solution to the target probability
maximization problem. It is a maximization of a law-invariant objective and
therefore it is path-independent. Denote it by $X_{T}^{*}:=f^{*}(S_{T}).$
Define $A_{0}=\{x\ |\ f^{*}(x)=0\}$, $A_{1}=\{x\ |\ f^{*}(x)=b\}$, $%
A_{2}=\{x\ |\ f^{*}(x)\in ]0,b[\}$ and $A_{3}=\{x\ |\ f^{*}(x)>b\}$. We show
that $\P (S_{T}\in A_{0}\cup A_{1})=1$ must hold. Assume $\P (S_{T}\in
A_{0}\cup A_{1})<1$ so that $\P (S_{T}\in A_{2}\cup A_{3})>0.$ Define 
\begin{equation*}
Y=\left\{ 
\begin{array}{cl}
f^{*}(S_{T}) & \hbox{for}\ S_{T}\in A_{0}\cup A_{1}, \\ 
0 & \hbox{for}\ S_{T}\in A_{2}, \\ 
b & \hbox{for}\ S_{T}\in A_{3}.%
\end{array}%
\right.
\end{equation*}%
Then we observe that $Y=f^{*}(S_{T})$ on $A_{0}\cup A_{1}$ and $%
Y<f^{*}(S_{T})$ on $A_{2}\cup A_{3}.$ Since $\P (S_{T}\in A_{2}\cup A_{3})>0$
also $\mathbb{Q}(S_{T}\in A_{2}\cup A_{3})>0$ because $\P $ and the risk
neutral probability $\mathbb{Q}$ are equivalent. Hence $c_{0}(Y)<W_{0}.$
Next we define $Z=b\mathds{1}_{S_{T}\in C}+Y\ $ where we have chosen $%
C\subseteq A_{2}\cup A_{0}$ such that $c_{0}(b\mathds{1}_{S_{T}\in
C})=W_{0}-c_{0}(Y).$ Since $\P (S_{T}\in C)>0$ one has that $\P (Z\geq b)>\P %
(Y\geq b)=\P (f^{*}(S_{T})\geq b).$ Hence $Z$ contradicts the optimality of $%
f^{*}(S_{T})$. 
Therefore $\P (S_{T}\in A_{0}\cup A_{1})=1.$ Hence $f^{*}(S_{T})$ can take
only the values 0 or $b$. Since it is increasing in $S_{T}$ almost surely
(by cost-efficiency) it must write as 
\begin{equation*}
f^{*}(S_{T})=b\mathds{1}_{S_{T}>a},
\end{equation*}%
where $a$ is chosen such that the budget constraint is satisfied. 
\hfill $\Box $

\subsection{Proof of Theorem \protect\ref{TO2bis}\label{proofTO2bis}}

The (random) target probability maximization problem is given as%
\begin{equation*}
\max_{X_{T}\geq 0,c_{0}(X_{T})=W_{0}}\P [X_{T}\geq B].
\end{equation*}%
Assume that there exists an optimal solution $X_{T}^{\ast }$ to this
optimization problem. There are three steps in the proof.

\begin{enumerate}
\item The optimal payoff is of the form $f(S_{T},B)$.

\item The optimal payoff is of the form $B\mathds{1}_{h(S_{T},B)\in A}$.

\item The optimal payoff is of the form $B\mathds{1}_{B\xi _{T}<\lambda ^{{%
\ast }}}$ for $\lambda ^{{\ast }}>0$.
\end{enumerate}

\noindent \textbf{Step 1:} We observe that $X_{T}^{\ast }$ has some joint
distribution $G$ with $B.$ Theorem \ref{main} implies there exists a twin $%
f(B,S_{T})$ such that $(f(B,S_{T}),B_{T})\sim (X_{T}^{\ast },B)\sim G$ and $%
c_{0}(f(B,S_{T}))=c_{0}(X_{T}^{\ast })=W_{0}$. Therefore $\P (f(B,S_{T})\geq
B)=\P (X_{T}^{\ast }\geq B)$ and $\P (f(B,S_{T})\geq 0)=\P (X_{T}^{\ast
}\geq 0)=1$. Thus $f(B,S_{T})$ is also an optimal solution.

\noindent \textbf{Step 2:} This is similar to the proof of Proposition \ref%
{BrP}, applied conditionally on $B.$ Define the sets $A_{0}=\{s,f(B,s)=0\}$, 
$A_{1}=\{s,f(B,s)=B\},$ then $\P (S_{T}\in A_{0}\cup A_{1}|B)=1$ and
therefore $\P (S_{T}\in A_{0}\cup A_{1})=1$. Thus there exists a set $A$ and
a function $h$ such that 
\begin{equation*}
f(B,S_{T})=B\mathds{1}_{h(S_{T},B)\in A}.
\end{equation*}

\noindent \textbf{Step 3:} Define $\lambda >0$ such that 
\begin{equation*}
\P (h(S_{T},B)\in A)=\P (B\xi _{T}<\lambda ).
\end{equation*}%
Observe that $\mathds{1}_{h(S_{T},B)\in A}$ and $\mathds{1}_{B\xi
_{T}<\lambda }$ have the same distribution and that in addition, $B\xi _{T}$
is anti-monotonic with $\mathds{1}_{B\xi _{T}<\lambda }$. Therefore by
applying Lemma \ref{lem1} one has that 
\begin{equation*}
c_{0}(B\mathds{1}_{B\xi _{T}<\lambda })=\mathbb{E}[B\xi _{T}\mathds{1}_{B\xi
_{T}<\lambda }]\leq \mathbb{E}[B\xi _{T}\mathds{1}_{h(S_{T},B)\in A}]
\end{equation*}%
and therefore the optimum must be of the form $B\mathds{1}_{B\xi
_{T}<\lambda ^{\ast }}$ where $\lambda ^{\ast }>\lambda $ is determined such
that $c_{0}(B\mathds{1}_{B\xi _{T}<\lambda ^{\ast }})=W_{0}.$%
\hfill $\Box $

\subsection{Proof of Theorem \protect\ref{TO2}\label{proofTO2}}

The target probability maximization problem is given by%
\begin{equation*}
\underset{\substack{ X_{T}\geq 0,\ c_{0}(X_{T})=W_{0}, \\ \mathcal{C}%
_{(X_T,A_T)}=C}}{\max }\P [X_{T}\geq b]
\end{equation*}%
Assume that there exists an optimal solution $X_{T}^{*}$ to this
optimization problem. There are three steps in the proof.

\begin{enumerate}
\item The optimal payoff is of the form $f(S_{T},A_{T})$.

\item The optimal payoff is of the form $b\mathds{1}_{h(S_{T},A_{T})\in A}$.

\item The optimal payoff is of the form $A_{T}\mathds{1}_{Z_{T}>\lambda ^{{*}
}}$ for $\lambda ^{{*} }>0$.
\end{enumerate}

\noindent \textbf{Step 1:} We observe that $X_{T}^{*}$ has some joint
distribution $G$ with $A_{T}.$ Theorem \ref{main} implies there exists a
twin $f(S_{T},A_{T})$ such that $(f(S_{T},A_{T}),A_{T})\sim
(X_{T}^{*},A_{T})\sim G$ and $c_{0}(f(S_{T},A_{T}))=c_{0}(X_{T}^{*})=W_{0}$.
Therefore $\P (f(S_{T},A_{T})\geq b)=\P (X_{T}^{*}\geq b)$ and $%
P(f(S_{T},A_{T})\geq 0)=\P (X_{T}^{*}\geq 0)=1$. Thus $f(S_{T},A_{T})$ is
also an optimal solution.

\noindent \textbf{Step 2:} This is similar to the proof of Proposition \ref%
{BrP}. Define the sets $A_{0}=\{(s,t),f(s,t)=0\}$, $A_{1}=\{(s,t),f(s,t)=b%
\}, $ then $\P (S_{T}\in A_{0}\cup A_{1})=1.$ Thus there exists a set $A$
and a function $h$ such that 
\begin{equation*}
f(S_{T},A_{T})=b\mathds{1}_{h(S_{T},A_{T})\in A}.
\end{equation*}

\noindent \textbf{Step 3:} Define $\lambda >0$ such that 
\begin{equation*}
\P (h(S_{T},A_{T})\in A)=\P (Z_{T}>\lambda ).
\end{equation*}%
Observe that $b\mathds{1}_{h(S_{T},A_{T})\in A}$ and $b\mathds{1}%
_{Z_{T}>\lambda }$ have the same joint distribution $G$ with distribution $%
A_{T}$. Therefore, Theorem \ref{main2} shows that, 
\begin{equation*}
c_{0}(b\mathds{1}_{Z_{T}>\lambda })\leq c_{0}(b\mathds{1}_{h(S_{T},A_{T})\in
A}).
\end{equation*}%
Hence, $b\mathds{1}_{Z_{T}>\lambda ^{\ast }}$ where $\lambda ^{\ast }$ such
that $c_{0}(b\mathds{1}_{Z_{T}>\lambda ^{\ast }})=W_{0}$ is the optimum. %
\hfill $\Box $

\subsection{Proof of formula (\protect\ref{exampleTPM})\label%
{proofexampleTPM}}

We know that $b\mathds{1}_{Z_{T}>\lambda ^{\ast }}$ where $\lambda ^{\ast }$
is such that $c_{0}(b\mathds{1}_{Z_{T}>\lambda ^{\ast }})=W_{0}$ is the
optimal solution. We find that 
\begin{eqnarray*}
Z_{T} &=&C_{1|S_{t}}^{-1}({F_{S_{T}|S_{t}}(S_{T})}) \\
&=&\Phi \left[ \sqrt{1-\rho ^{2}}\left( \frac{\ln \left( \frac{S_{T}}{St}%
\right) -\left( \mu -\frac{\sigma ^{2}}{2}\right) (T-t)}{\sigma \sqrt{T-t}}%
\right) +\rho \left( \frac{\ln \left( \frac{S_{t}}{S_{0}}\right) -\left( \mu
-\frac{\sigma ^{2}}{2}\right) t}{\sigma \sqrt{t}}\right) \right] .
\end{eqnarray*}%
It is then straightforward that $X_{T}^{\ast }=b\mathds{1}_{\left\{
S_{t}^{\alpha }S_{T}>\lambda ^{\ast }\right\} }$ is the optimal solution,
with $\alpha $ and $\lambda $ given by 
\begin{eqnarray*}
\alpha &=&\sqrt{\frac{T-t}{t(1-\rho ^{2})}}\rho -1 \\
\lambda &=&S_{0}^{1+\alpha }\exp \left( \left( r-\frac{\sigma ^{2}}{2}%
\right) (\alpha t+T)-\sigma \sqrt{(\alpha +1)^{2}t+(T-t)}\Phi ^{-1}\left( 
\frac{W_{0}e^{rT}}{b}\right) \right) .
\end{eqnarray*}%
{\ }\hfill $\Box $

\newpage \singlespacing

\end{document}